\documentclass[5p, times]{elsarticle}
\usepackage{amsfonts}
\usepackage{amsmath}
\usepackage{algorithm,algorithmicx,algpseudocode}
\usepackage{subfig}
\usepackage{multirow}
\usepackage{url}
\usepackage{graphicx}
\usepackage{caption}
\usepackage{booktabs}
\usepackage{balance}

\usepackage{enumitem}

\usepackage{array}
\usepackage{color}
\usepackage{colortbl}

\DeclareGraphicsExtensions{.pdf,.png,.jpg,.eps}

\algnewcommand\algorithmicforeach{\textbf{for each}}
\algdef{S}[FOR]{ForEach}[1]{\algorithmicforeach\ #1\ \algorithmicdo}
\algnewcommand{\algorithmicand}{\textbf{ and }}
\algnewcommand{\algorithmicor}{\textbf{ or }}
\algnewcommand{\algorithmicnot}{\textbf{ not }}
\algnewcommand{\OR}{\algorithmicor}
\algnewcommand{\AND}{\algorithmicand}
\algnewcommand{\NOT}{\algorithmicnot}

\usepackage[]{changes}
\definechangesauthor[color=red]{AP}
\definechangesauthor[color=blue]{EB}
\definechangesauthor[color=green]{RB}

\usepackage{soul}

\sethlcolor{yellow}

\usepackage[]{todonotes}

\makeatletter
\def\ps@pprintTitle{%
  \let\@oddhead\@empty
  \let\@evenhead\@empty
  \let\@oddfoot\@empty
  \let\@evenfoot\@oddfoot
}
\makeatother

\begin{document}
\begin{frontmatter}

\title{Reliable data delivery in ICN-IoT environments}

\author[cnr]{Eleonora Borgia\corref{cor1}} \ead{eleonora.borgia@iit.cnr.it}
\author[cnr]{Raffaele Bruno} \ead{raffaele.bruno@iit.cnr.it}
\author[cnr]{Andrea Passarella} \ead{andrea.passarella@iit.cnr.it}
\cortext[cor1]{Corresponding author}
\address[cnr]{Institute for Informatics and Telematics, National
Research Council. Via G. Moruzzi 1, 56124 Pisa, Italy.}

\begin{abstract}
In an IoT environment, which is characterized by a multitude of interconnected smart devices with sensing and computational capabilities, many applications are (i) content-based, that is, they are only interested in finding a given type of content rather than the location of data, and (ii) contextualized, that is, the content is generated and consumed in the proximity of the user. 
In this context, the Information-Centric Networking (ICN) paradigm is an appealing model for efficiently retrieving application data, and the service decentralization towards the network edge helps to reduce the core network load being the data produced by IoT devices mainly confined in the area where they are generated. 
MobCCN is an ICN-based data delivery protocol that we designed for operating efficiently in such context \cite{Borgia:2016:MCP:2979683.2979695, BORGIA201881}, where static and mobile IoT devices are enriched with ICN functions. Specifically, MobCCN leverages an efficient routing and forwarding protocol, exploiting opportunistic contacts among IoT mobile devices, to fill the Forwarding Interest Base (FIB) tables so as to correctly forward Interest packets towards the intended data producers. 
In this paper, we aim to enhance the reliability of MobCCN by exploring different retransmissions mechanisms, such as retransmissions based on number of duplicate Interests that are received for the same requested content, periodic retransmissions, single path versus disjoint multi-path forwarding, hysteresis mechanism and combinations of them. 
Extensive simulation results show that, among the analysed MobCCN variants, the one that implements both periodic retransmissions and a hysteresis-based retransmission process ensures the highest delivery rates (up to 95\%) and the shortest network path, with a very limited traffic overhead due to retransmissions. 
\end{abstract}


\end{frontmatter}

\section{Introduction}
\label{sec:intro}
\begin{figure}[t]
 	\centering
	\includegraphics[trim={0, 0, 0, 0}, clip, scale=0.3]{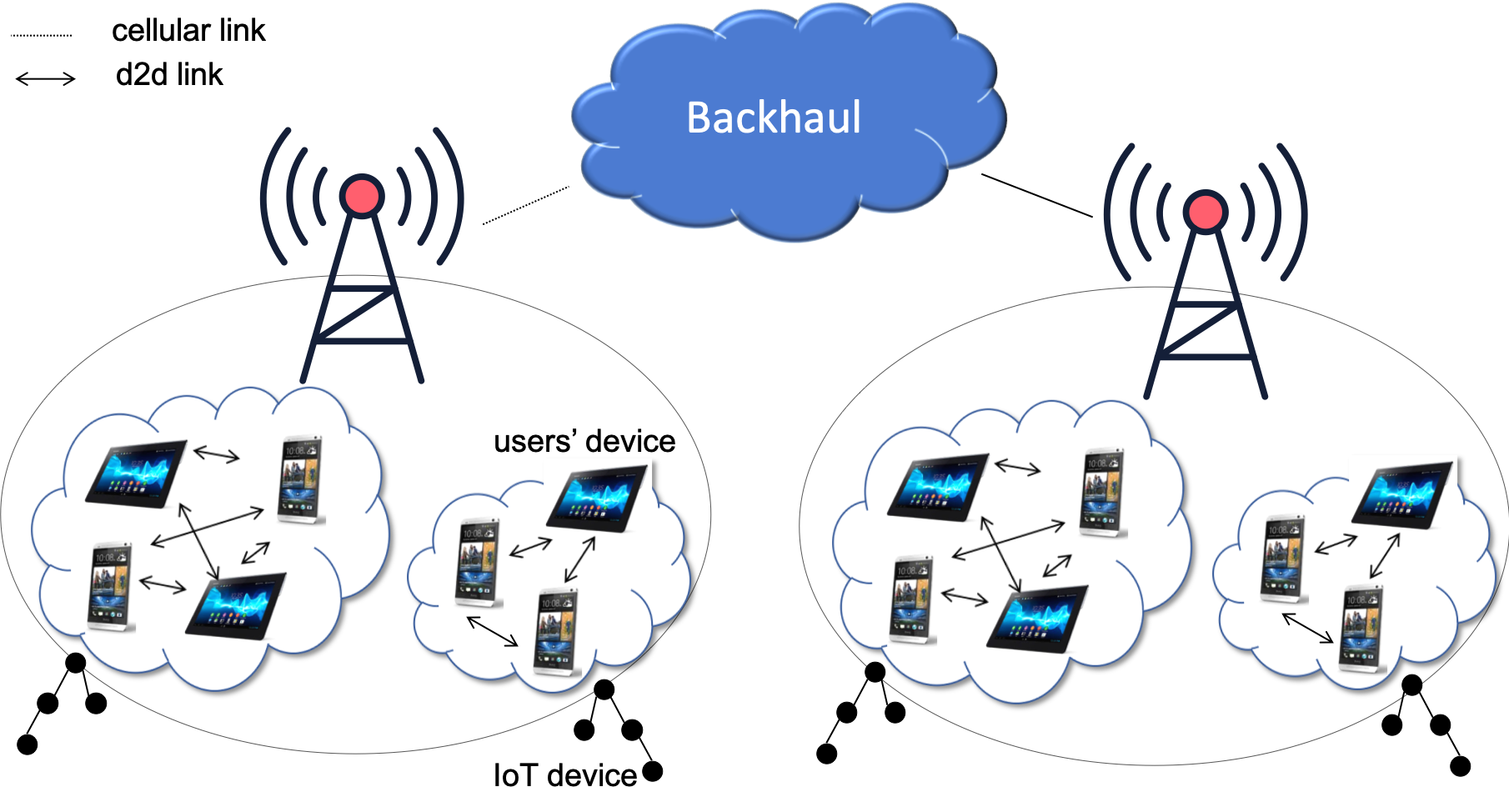}
	\vspace{-0.3cm}
 	\caption{Illustrative example of the reference network scenario considered in this study.}
	\vspace{-0.3cm}
 	\label{fig:ref-arch}
	\vspace{-0.3cm}
 \end{figure}

\noindent
The \emph{Internet of Things (IoT)} refers to the paradigm where a massive number of inter-connected devices with sensing and computational capabilities, uniquely addressable, forms a dynamic network to communicate with each other or with the Internet without human intervention \cite{Borgia20141}. IoT devices are extremely heterogeneous in terms of resource capabilities, lifespan and communication technologies. On the one end of the spectrum, there are static, tiny and battery-powered devices with limited computational and memory capabilities, which are embedded with sensors and actuators. These IoT devices are typically used to collect and exchange sensed data about the state of physical objects, processes and environments, and to react to this information by performing appropriate control actions. On the other end of the spectrum, the physical environment is also sprinkled with more powerful IoT devices, e.g., smartphones carried by mobile users, or vehicle-based sensing platforms, which can produce, collect, request, consume data through context-aware applications. Previous research studies have proposed to leverage these less resource-constrained mobile devices to act as ``data mules'' so as to facilitate the collection and distribution to other nodes of the data generated by the tiny sensor nodes \cite{2011_TOSN_data_mule,Borgia2016p,2018_JSYST_data_mule}. Recently, this architectural design has gained further popularity thanks to new 4G/5G capabilities, and in particular D2D communications and Proximity Services (ProSe), which provide the enabling mechanisms to discover other mobile devices in close proximity and to communicate with them directly \cite{Lin2014,Jameel2018}. Motivated by these considerations, we envision a scenario where at the edge of the network there are regions where IoT data is produced by local static IoT devices, and this data is collected and distributed to other devices in the overall area by leveraging an intermediate layer of mobile users' devices, as shown in Figure \ref{fig:ref-arch}. Typical application scenarios for this framework can be identified in the smart-city context. For instance, localized information about discounts or coupon deals from nearest shops in a mall can be forwarded using D2D communications between users' personal devices to the people walking in the area. Similarly, images captured by camera-based sensors that are installed in city hot spots can forwarded to the interested users without using expensive cellular communications. Note that these applications are typically delay tolerant and not critical, namely minimal latency is not a critical factor. Furthermore, the intrinsic data redundancy (e.g., multiple cameras taking similar images of a scene) makes acceptable to loose some of the transmitted messages.

From the above discussion, it should be clear that this study specifically targets the emerging category of \emph{content-centric} IoT applications rather than host-centric ones. In such type of applications, the goal of the communication process is to search for a specific content (mainly generated in the proximity of the user), and not to identify the location of the device that originated or currently stores that content. It is important to remind that the \emph{Information-Centric Networking} (ICN) architecture \cite{6563278} was originally proposed to support content-centric applications in the Internet, such as file sharing and media streaming. However, the ICN paradigm has been also proposed as an alternative networking architecture for IoT \cite{6882665,Baccelli:2014:ICN:2660129.2660144}. The basic ICN design principle is to adopt content naming, meaning that each content is assigned a unique and location-independent name. The content retrieval process then follows a receiver-driven approach, as content segments (or chunks) are exchanged using a request/response model. Additional appealing ICN features are in-networking caching, request aggregation, mobility support, and content security \cite{6563278}.

As the integration of ICN mechanisms into IoT environments has a strong potential to fulfil the requirements of content-centric IoT applications, in this work we investigate how to implement an ICN-based framework in the scenario illustrated in Figure~\ref{fig:ref-arch}, to provide reliable and scalable IoT data collection and distribution by leveraging users' mobile devices. In our previous work we have already proposed MobCCN, a routing and forwarding protocol that is compliant with ICN paradigms (namely CCN \cite{Jacobson:2009:NNC:1658939.1658941} and NDN \cite{Zhang:2014:NDN:2656877.2656887}), and leverages D2D communications to allows users' mobile devices to opportunistically exchanged data collected by nearby static IoT devices. To perform precise forwarding decisions, MobCCN exploits the concept of utility of a node towards a given content, i.e., a measurement of how frequently a node meets other nodes that store the requested content in their local caches. These utility values are exchanged during opportunistic contacts among mobile nodes, to create a dynamic gradient-based content-dissemination graph used to redirect content requests and retrieve the content efficiently. 
We presented the main idea of MobCCN in~\cite{Borgia:2016:MCP:2979683.2979695} and its performance in~\cite{BORGIA201881}, where we have proved that, although additional signalling traffic is needed to build more efficient forwarding tables by computing and exchanging the utility value for each type of content available in the network, this approach ensures a more precise detection of either the relevant content producers or the nodes that cache the relevant content. This yields a remarkable reduction of total network traffic and cache usage when compared with an epidemic-based routing protocol with a small degradation of the delivery rate in most cases, and an increase of end-to-end delays. 

The retransmission-based loss recovery mechanisms typically used by CCN, which are inherited in MobCCN, are known to be not very efficient in lossy or mobile networks \cite{2014_icn_retr}. The goal of this study is to shed some light on the performance improvements that could be achieved if more sophisticated retransmission mechanisms are used in MobCCN, and the consequential trade-offs between increased reliability (i.e., higher delivery rate), and delays or network traffic overheads. Specifically, the standard CCN retransmission scheme simply retransmits pending Interest packets periodically or when a duplicate one is received, using the reverse path traversed by the interest packet. In this work we explore alternative retransmission approaches. Specifically, we consider schemes that leverage \emph{path diversity} (i.e., the packet loss processes are expected to operate independently for different paths) by sending retransmissions over a different path than the reverse path. Note that multi-path routing is also beneficial to distribute more evenly the traffic load due to retransmissions. In addition, we also investigate adaptive scheme to set up the periodic retransmission timer, which leverages \emph{utility hysteresis} to control the retransmission frequency. We have integrated the proposed retransmission strategies into the MobCCN prototype, and evaluated the advantages and disadvantages of each of them under a variety of scenarios using the OMNeT++ network simulator\footnote{http://www.omnetpp.org/}. Simulation results show that the hysteresis-based approach, called MobCCN\_AH, achieves the best performance, with delivery rates in the range of [85-95]\%, about 15-20\% higher than the ones obtained by basic MobCCN, when sufficient bandwidth is available for the retransmissions. On the contrary, in bandwidth-limited scenarios the performance of the different retransmission policies tends to converge.

The rest of the paper is organised as follows. Section~\ref{sec:related} overviews the most related research work. Section~\ref{sec:basic-features} briefly presents the basic MobCCN protocol. In Section~\ref{sec:mobccn-enhancements} we describe the proposed retransmission strategies. Section~\ref{sec:simu-setup} introduces the simulation setup.  
In Section~\ref{sec:perf-eva} we present the comprehensive performance comparison. Finally, Section~\ref{sec:conclusions} draws the main conclusions of the paper.

\section{Related work}
\label{sec:related}
\noindent

In the recent years, an increasing number of research studies have been dedicated to investigate critical aspects and challenges for the integration of ICN mechanisms into IoT networks \cite{7437030, 8478349, NOUR201995, Djama:2020}. 

One line of research that is well investigated focuses on the problem of efficient data retrieval, which consists in specifying the routing and forwarding processes. Proactive link-state routing solutions, where nodes periodically exchange information to fill the forwarding tables, have been typically proposed for network scenarios with static nodes (see for example OSPFN \cite{OSPFN:2012} or NLSR \cite{NLSR:Hoque:2013}). On the contrary, routing protocols for ICN-IoT environments with mobile nodes are less investigated. DABBER \cite{Mendes:2018}, for instance, supports opportunistic communications. It relies on node context information such as availability and centrality of adjacent nodes, as well as of the availability of different data sources to drive forwarding of Interests. 
When an underlying routing protocol populating FIBs is missing, the forwarding plane acts as a reactive routing protocol by exploiting on-demand or controlled flooding to decide the best next-hop forwarder \cite{LIU:2017}. To further refine forwarding decisions, some solutions use the context of the requested content and the tolerance to inaccuracies in data defined for the application requesting it \cite{J.S.M.:2016}, or the network performance measurements \cite{YI:2013}, or a mix of context information and optimal selection of the forwarding times \cite{Borrego:2020}. As an alternative to name-based flooding, location-based solutions that exploit the geographic coordinates of nodes are also investigated, primarily in the context of vehicular networks \cite{Grassi:2015} and recently in IoT systems \cite{Enguehard:2018}. To improve the scalability of forwarding tables, solutions based on bloom filters are proposed in \cite{Rodrigues:2018, Antikainen:WTS16, Tapolcai:2015}. There exists also a class of routing protocols for ICN-IoT environments specifically designed for push-based traffic, such as that produced in case of alarms or status changes. These schemes allow data producers to distribute data to subscribed consumers even without explicit and continuous requests \cite{Amadeo:2014:push, Moll:2017}.

The mobility of nodes increases the complexity level of designing an efficient data retrieval. 
Mobile devices can be content consumers, or content producers, but they can also play the role of consumers and producers at the same time. Basically, ICN networks support consumer mobility by design. In contrast with IP-networks, where an update of the address information is needed each time a mobile node changes the network as happens in Mobile IP, the native mobility ICN support consists just in reissuing the content request for those lost contents. However, such solution is not always sufficient, especially in the IoT context, but more complex schemes are required such as rendezvous-based schemes \cite{Anastasiades-2014}, or synchronisation of the subscription table during consumers' movements \cite{Nour:2017}. On the contrary, providers' mobility is more challenging because it is necessary to maintain track of the providers' locations in order to guarantee the session maintenance. Among the several classes of proposed solutions, multi-path based schemes simultaneously transmit multiple copies of content requests to potential locations where the providers may move \cite{Ravindran:2012}. The use of caching also contributes to mitigate movements of both consumers \cite{Vasilakos:2012} and producers \cite{Araujo:2019}, as well as improving system performance, especially in terms of experienced delay.

To the best of our knowledge, the design of reliability mechanisms to improve the robustness of an ICN-based data delivery process has not been sufficiently investigated. Most of the existing solutions assign the execution of the proposed reliability mechanisms to the consumers, which are in charge of simply re-issuing Interest packets for those Data packets that were never received.  
Retransmissions are usually triggered by the application layer when a specific retransmission timeout (RTO) for a content expires. Almost all solutions focus exclusively on the estimation of such timeout, for example by including a mechanism similar to TCP EWMA, or by excluding any RTO updates in case of Data reception due to retransmissions \cite{Carofiglio:2012, Amadeo:2014:InfocomWkshps}, or estimating different timeout values for each content producer \cite{Anastasiades:2015}. None of the aforementioned studies attempts to improve the retransmission mechanisms at the network level nor do they explore the effectiveness of retransmission mechanisms that are not based on timeouts, which is one of the main objectives of this work. In our previous paper \cite{BORGIA201881} we have already proved that, when applying the standard retransmission mechanism adopted by the vanilla ICN, there are cases in which the delivery rate does not improve but only a traffic overhead increase is observed. An additional issue that we have also observed is the occurrence of network loops for Interest packets. Thus, further investigations are needed to improve the network-level retransmission mechanism in  ICN systems. A recent and orthogonal direction is the use of network coding in ICN networks to increase the robustness to content delivery \cite{draft-matsuzono-nwcrg-nwc-ccn-reqs}. Some solutions are presented in \cite{Saltarin:2016, Xu:2018}. However, both works assume communications over wired networks to which devices with multiple (wireless) interfaces try to access, therefore a very different environment with the one considered in this paper.   
 
Finally, several prototypes of ICN-based solutions for IoT networks have been developed using software platforms and operating systems that were specifically designed for resource-constrained devices, such as CCN-lite~\cite{ccn-lite}, RIoT~\cite{BaccelliRIOT}, and Contiki~\cite{Dunkels2004}. Large-scale real-world test-beds have been used to analyse the performances of these prototypes from different perspectives, such as resource (e.g., memory, cpu, energy) consumption~\cite{Baccelli2014, Ahlgren2016}, comparison between  single-hop and multi-hop network typologies~\cite{Baccelli2014,Cenk2018lcn}, and comparison of ICN-based architectures against IP-based ones~\cite{Baccelli2014,Cenk2018icn}. However, most of these experimental studies consider only static network scenarios.

\section{Background: the MobCCN protocol}
\label{sec:basic-features}
\noindent
\begin{figure}[ht]
\centering
    \subfloat[t][\label{fig:MobCCNprotocol:routing}]{%
	\centering
      \includegraphics[trim={0cm 0cm 0cm 0cm},clip,width=0.45\textwidth]{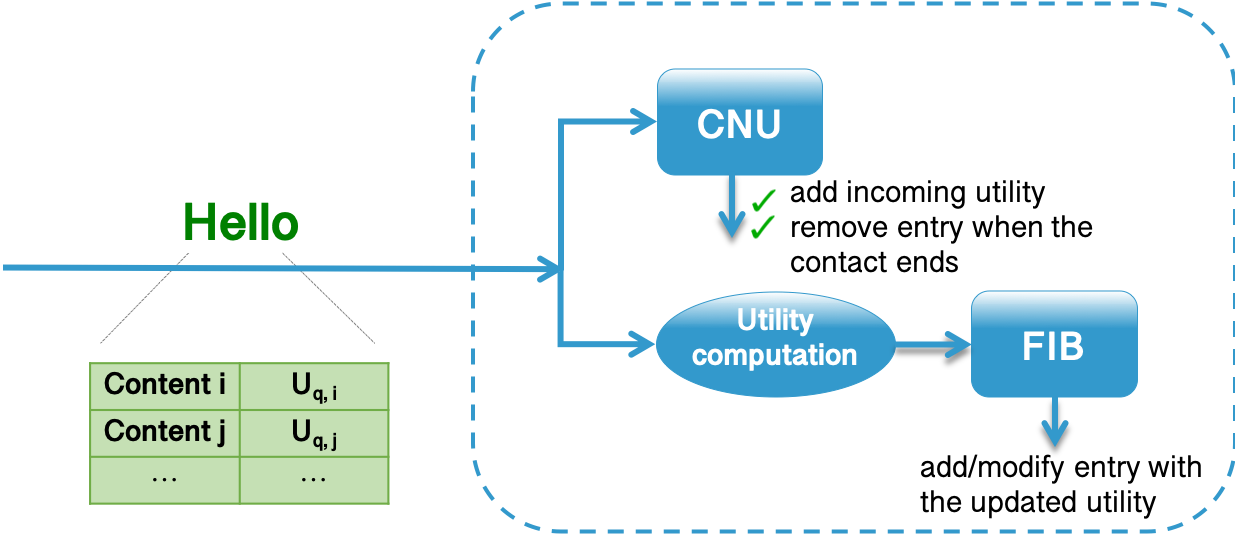}
    }
\\
%
     \subfloat[\label{fig:MobCCNprotocol:forwarding}]{%
      \includegraphics[trim={0cm 0cm 0cm 0cm},clip,width=0.48\textwidth]{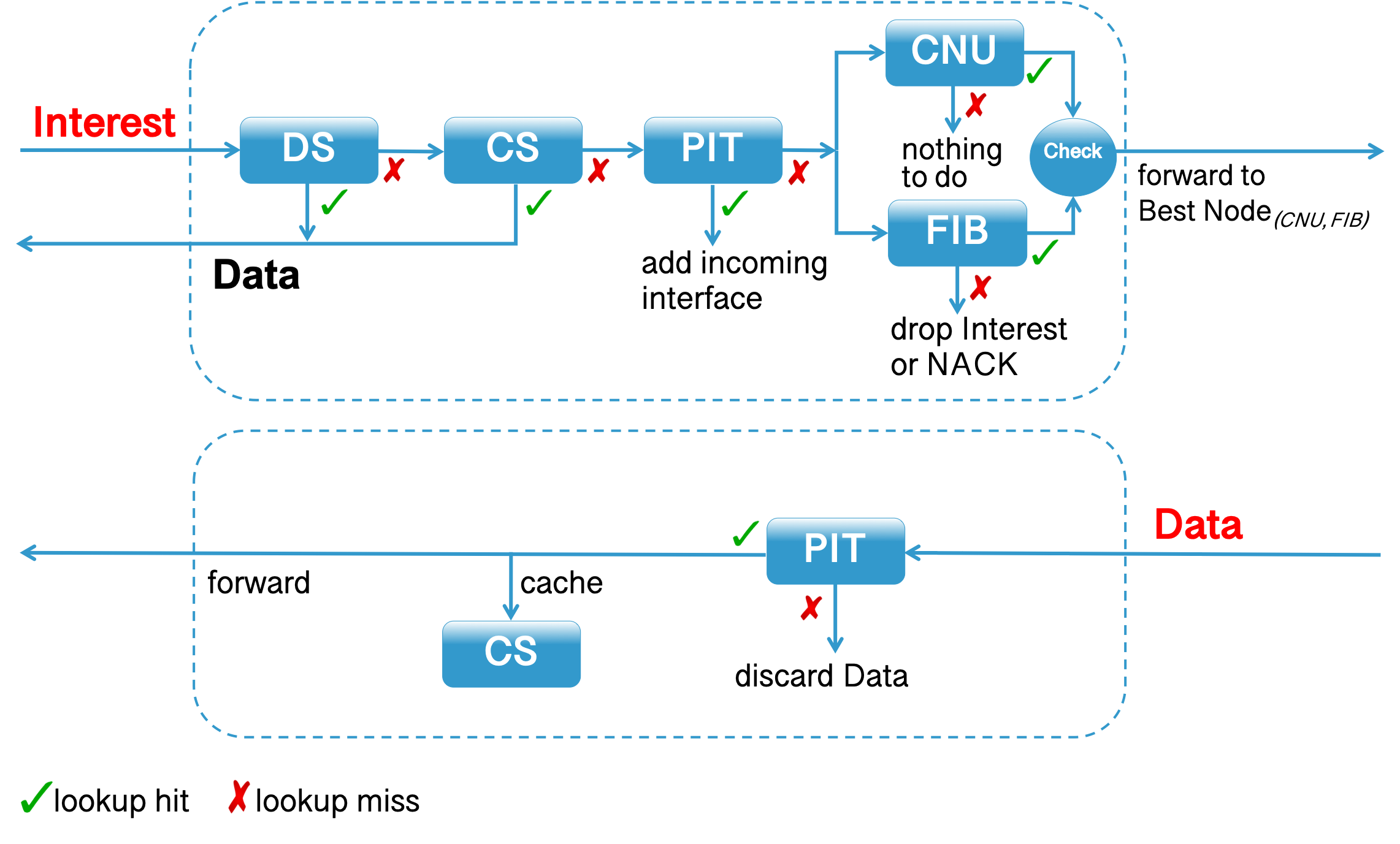}

      }
\vspace{-0.1cm}
\caption{MobCCN protocol: routing phase (\ref{fig:MobCCNprotocol:routing}), and forwarding phase (\ref{fig:MobCCNprotocol:forwarding}).}	\label{fig:MobCCNprotocol:}
\vspace{-0.3cm}
\end{figure}
For the sake of clarity, in this section we overview the general design principles and the basic mechanisms of  MobCCN protocol. We refer to \cite{Borgia:2016:MCP:2979683.2979695} and to \cite{BORGIA201881} for a more detailed description. 

As explained in Section \ref{sec:intro}, the design rationale of MobCCN is to use an ICN-based communication paradigm to build a data management layer that is made of mobile devices using D2D communications to exchange messages. This data management platform is then used to collect and distribute data generated by nearby sensor nodes and other mobile devices. MobCCN inherits all the basic mechanisms and concepts of ICN solutions, and in particular the ones of the CCN/NDN architecture. Specifically, in MobCCN, Interest packets are used to request contents, Data packets carry the requested contents, the Content Store (CS) is used for storing the content received by a node, the Pending Interest Table (PIT) takes note of Interest packets arrived at one node but not yet satisfied, and the Forwarding Interest Table (FIB) serves as the routing table of Interest packets, similarly to the packet routing table of a legacy Internet node. In MobCCN a data store, called DS, is used not only to maintain the data that is produced by the node itself, but also to store the contents that are retrieved by the sensor nodes that are encountered. The key concept behind MobCCN routing process is the \emph{utility value} of a node $p$ for a specific content $i$ ($U_{p, i}$), which is a function of both the frequency node $p$ encounters the node that generated content $i$ (\emph{direct utility} component) and the frequency node $p$ encounters other nodes that have earlier ``bumped'' into that content, i.e., which have non-null utility towards content $i$ (\emph{indirect utility} component). Formally, the direct utility of a node $p$ towards a content $i$ is defined as: 
\begin{equation}
  U_{p,i}^{(d)} = \frac{1}{ICT(p,i)} \; 
  \label{eq:direct-utility}
\end{equation}

where $ICT(p,i)$ is the inter-contact time between $p$ and any node that generated content $i$.  The indirect utility is defined as:
\begin{equation}
  U_{p,q,i}^{(ind)} = \frac{1}{ \frac{1} {U_{q,i}} + ICT(p,q)}  \; , 
  \label{eq:indirect-utility}
\end{equation}

where $q$ is any node encountered by node $p$, which does not store the content $i$, $U_{q,i}$ is the current utility of node $q$ towards content $i$, and $ICT(p,q)$ is the inter-contact time between $p$ and $q$. Direct and indirect utilities are then merged together to provide the global utility value of the node $p$ for the content $i$ as follows:
\begin{equation}
  U_{p,i} = \max_{j \in \mathcal{N}_p} \left\{U_{p,i}^{(d)},U_{p,j,i}^{(ind)}\right\} \; ,
  \label{eq:utility}
\end{equation}
where $\mathcal{N}_p$ is the set of nodes that $p$ met.
The utility value $U_{p,i}$, computed and updated at each contact by exchanging \emph{Hello} packets, is used to create a \emph{utility-gradient-based information-centric graph} during the routing phase, and to forward the Interest packets during the forwarding phase.

MobCCN uses the \emph{non-permanent} in-network caching policy of ICN, namely each node stores into its CS a copy of the content items it receives, but only for the amount of time that is necessary to serve all pending requests in the PIT. Therefore, in MobCCN a node does not announce a direct utility value for the contents that are added to its CS as they are only temporarily stored and the time they are maintained into the case is variable. In fact, using these copies of the content in the direct utility computation may lead to inconsistent routing decisions in other nodes. During the routing phase (see Figure \ref{fig:MobCCNprotocol:routing}) when two nodes meet, each of them broadcasts a Hello packet to advertise the utilities of the contents it knows, as well as the identifiers of the nodes with the highest utility values for those contents. Then, for each content $i$ advertised in the received Hello packet the tagged node $p$: \emph{1)} creates a temporary entry in the Current Neighbours Utilities (CNU) table for content $i$ by storing the utility received by the neighbour\footnote{This entry is then removed upon the expiration of contact.}, \emph{2)} computes the new utility value using eq. \ref{eq:direct-utility} or eq. \ref{eq:indirect-utility} depending on whether the encountered node is a producer or not for content $i$, and \emph{3)} creates a new (or updates the corresponding) entry in the FIB table for content $i$ by storing the interface (or ``face"\footnote{As we adopted the same CCN terminology, we use node/face/interface interchangeably in the text.}) from which the Hello packet came and the utility value computed earlier. In this way, the FIB table is populated with the information required to build a utility-gradient-based information-centric graph for all the contents available in the network.
 \begin{figure}[b]
 	\centering
	\includegraphics[trim={0, 0, 0, 0}, clip, scale=0.35]{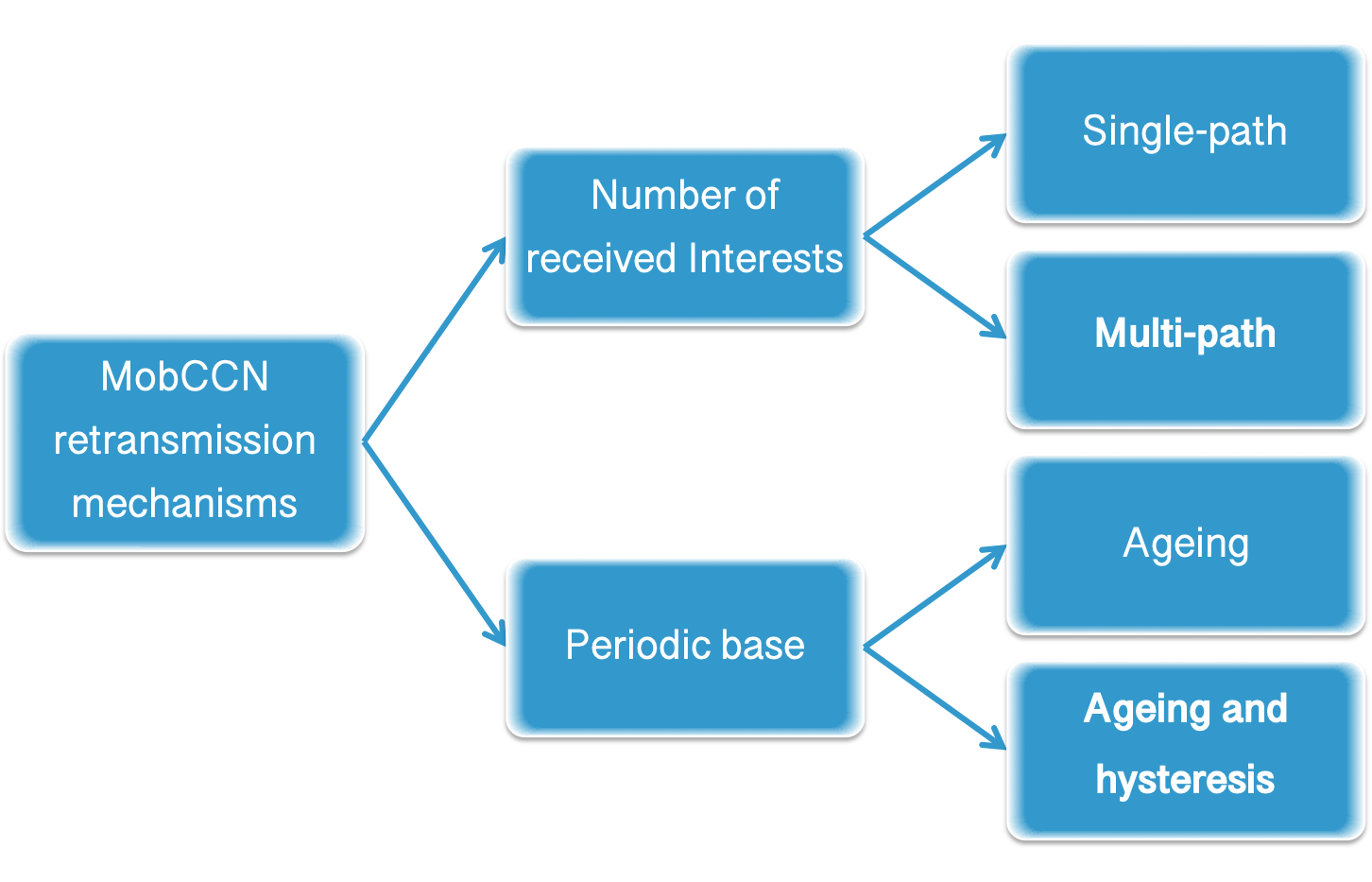}
	\vspace{-0.3cm}
 	\caption{Schematic representation of the MobCCN retransmission mechanisms.}
 	\label{fig:retx-scheme}
	\vspace{-0.3cm}
 \end{figure}

During the forwarding phase (see Figure \ref{fig:MobCCNprotocol:forwarding}), when a node $p$ receives an Interest packet for the content $i$, first it checks if the requested content is stored locally in CS and, if the case, it forwards the Data packet to the node from which the request arrived. If the content is not stored in the CS, it generates or updates the corresponding PIT entry. These two operations are performed exactly as in the original CCN. However, unlike CCN, the utility-gradient-based information-centric graph created during the routing phase is used to forward the Interest packet. Specifically, node $p$ identifies the node with the highest utility towards content $i$, searching both its current neighbours and other nodes from which it collected routing information in previous contacts. More precisely, MobCCN queries the CNU table to find the neighbour with the highest utility towards content $i$; and the FIB table to find the best candidate forwarders for that Interest packet. Only if the two nodes are the same, the Interest packet is forwarded immediately. Otherwise, the Interest packet is stored at node $p$, which repeats this selection procedure whenever the set of current neighbours changes. Forwarding the Interest packet following opportunistically the positive gradient ensures that the required content is discovered efficiently. Once the content is found, it is encapsulated in a Data packet and forwarded to the requester following the standard CCN mechanism that exploits the reverse-path established at each node.
%

%
\section{MobCCN reliability enhancements}
\label{sec:mobccn-enhancements}
\noindent
In the following we describe the retransmission mechanisms that we added to MobCCN in order to improve its reliability. Specifically, Section \ref{sub:retransmission-Interest-number} describes the two mechanisms that leverage the number of received Interests to trigger the retransmission of a content, while Section \ref{sub:retransmission-timeout} focuses on mechanisms based on timer expiration. It is worth noting that one of the two mechanisms of each category is inherited from the ones adopted by the vanilla CCN, and they are used as benchmarks for those that are proposed by this study, and which represent the original contribution of this paper. For the sake of presentation clarity, Figure \ref{fig:retx-scheme} shows a schematic classification of the retransmission mechanisms that we discuss in the following, where we mark our original contributions in bold.

\setlength{\textfloatsep}{0pt}
\begin{algorithm}[t]
\scriptsize
		\begin{algorithmic}[1] 
			\Require $E_p$ \Comment{Set of current neighbours of node $p$}
			\Statex
			\State $i \gets Interest.\mathtt{Name}()$
			\If {$\mathrm{Data} \gets \mathrm{CS}.\mathtt{Find}(i)$}
				\State \Return Data 
			\EndIf
			\If {$\mathrm{PIT}.\mathtt{Find}(i)$}
				\State $\mathrm{PIT}.\mathtt{AddFace}(i,q)$
				\If {$\mathrm{PIT}.\mathtt{numberOfFace}(i) > \emph{ReTX\textsubscript{threshold}}$} \Comment{old Interest, nevertheless \\ \hspace{6cm}propagated with R1}
					\State \Call {$\mathrm{R1}$}{$\mathrm{i, Interest}$}
					
			\EndIf
		    \Else
		    	\State $\mathrm{PIT}.\mathtt{CreateFace}(i,q)$
				\If {$\mathrm{FIB}.\mathtt{Find}(i)$}
					\State find the node $j$ with the highest utility for content type $i$ in FIB and CNU tables
					\If {$j \in E_p$}
						\State forward Interest packet to node $j$
					\Else
						\State store Interest packet waiting for the next contact  
				\EndIf
				\Else
					\State drop  Interest packet
				\EndIf
			\EndIf
			
			\Statex
			\Procedure{R1}{\emph{$i, $Interest}}
				\If {$\mathrm{FIB}.\mathtt{Find}(i)$}
					\State find the node $j$ with the highest utility for content type $i$ in FIB and CNU tables
					\If {$j \in E_p$}
						\State forward Interest packet to node $j$
					\Else
						\State wait for the next contact  
				\EndIf
				\Else
					\State drop  Interest packet
				\EndIf

			\EndProcedure
			
		\end{algorithmic}
\caption{\small{MobCCN\_R1: processing of Interest packets received by node $p$}}\label{alg:MobCCN_R1_interest_processing}
\end{algorithm}
\subsection{Retransmission mechanisms based on the number of received requests}
\label{sub:retransmission-Interest-number}
\noindent
In the following sub-section we describe the two solutions that differ in the way they route the pending Interest packets. 
%
\subsubsection{Single-path mechanism}
\label{sub:MobCCN-R1}
\noindent
First of all, we analyse the standard retransmission mechanism adopted by the vanilla CCN in which a retransmission occurs according to the number of the Interest packets for the same content received by a node. 
Algorithm \ref{alg:MobCCN_R1_interest_processing} describes the operations of this scheme executed when Interest packets arrive at one node.
More specifically, if a node $p$ receives an Interest packet that it cannot satisfy (i.e., the content is not stored in the CS), it generates a PIT entry and forwards the Interest to the node with the highest utility for that content among those in FIB and CNU tables (line 10-21). If $p$ receives a second Interest packet for the same content $i$ of which the corresponding Data message is still missing, it adds the incoming ``face" to the corresponding PIT entry (line 6). If the number of received Interests for the same content exceeds a certain retransmission threshold (i.e., ReTX\textsubscript{threshold} at line 7), $p$ retransmits the first received Interest towards the node with the highest utility for that content (line 23-34). It is important to observe that if the selected forwarder is not currently in proximity of node $p$ (i.e., the forwarder is not a neighbour of node $p$), the retransmission is delayed till the next contact. Intuitively, if the $ReTX_{threshold}$ is set equal to 0 then each duplicate reception of an Interest packet triggers a retransmission.  
Obviously, the smaller the $ReTX_{threshold}$ and the higher the retransmission traffic that is generated. Nevertheless, we expect that a trade-off exists between the additional network congestion caused by the transmitted packets and the gain in number of data requests that can be successfully satisfied.
In the rest of the paper, we refer to this solution as ``\emph{MobCCN\_R1}''.

\setlength{\textfloatsep}{0pt}
\begin{algorithm}[t]
\scriptsize
		\begin{algorithmic}[1] 
			\Require $E_p$ \Comment{Set of current neighbours of node $p$}
			\Statex
			\State $i \gets Interest.\mathtt{Name}()$
			\If {$\mathrm{Data} \gets \mathrm{CS}.\mathtt{Find}(i)$}
				\State \Return Data 
			\EndIf
			\If {$\mathrm{PIT}.\mathtt{Find}(i)$}
				\State $\mathrm{PIT}.\mathtt{AddFace}(i,q)$
				\If {$\mathrm{PIT}.\mathtt{numberOfFace}(i) > \emph{ReTX\textsubscript{threshold}}$} \Comment{old Interest, nevertheless \\ \hspace{6cm}propagated with R2}
					\State \Call {$\mathrm{R2}$}{$\mathrm{i, Interest}$}
					
				\EndIf
		    \Else
		    	\State $\mathrm{PIT}.\mathtt{CreateFace}(i,q)$
				\If {$\mathrm{FIB}.\mathtt{Find}(i)$}
					\State find the node $j$ with the highest utility for content type $i$ in FIB and CNU tables
					\If {$j \in E_p$}
						\State forward Interest packet to node $j$
					\Else
						\State store Interest packet waiting for the next contact  
				\EndIf
				\Else
					\State drop  Interest packet
				\EndIf
			\EndIf
			
			\Statex
			\Procedure{R2}{\emph{$i, $Interest}}
				\If {$\mathrm{FIB}.\mathtt{Find}(i)$}
					\State find the node $j$ with the second highest utility for content type $i$ in FIB and\\ \hspace{0.7cm}CNU tables
					\If {$j \in E_p$}
						\State forward Interest packet to node $j$
					\Else
						\State wait for the next contact  
				\EndIf
				\Else
					\State drop  Interest packet
				\EndIf

			\EndProcedure
		\end{algorithmic}
\caption{\small{MobCCN\_R2: processing of Interest packets received by node $p$}}\label{alg:MobCCN_R2_interest_processing}
\end{algorithm}
\setlength{\textfloatsep}{0pt}
\begin{algorithm}[ht]
\scriptsize
		\begin{algorithmic}[1] 
				\Require $E_p$ \Comment{Set of current neighbours of node $p$}

		\Statex
			
		\ForEach{$e\in\mathit{PIT}$}
			\If {$ \NOT \mathrm{PIT}.\mathtt{Expired}(e)  $}
    				\State $i \gets \mathrm{PIT}.\mathtt{Name}(e)$
    				\If {$\mathrm{FIB}.\mathtt{Find}(i)$}
					\State find the node $j$ with the highest utility for content type $i$ in FIB and \\ \hspace{1.1cm}CNU tables
					\If {$j \in E_p$}
						\State forward Interest packet to node $j$
					\Else
						\State wait for the next contact  
				\EndIf
				\Else
					\State drop  Interest packet
				\EndIf
			\Else
			    \State drop Interest packet
				\State PIT.remove(e)
			\EndIf
		
 		 \EndFor
			
		\end{algorithmic}
\caption{\small{MobCCN\_A: procedure at node $p$ at the expiration of the ageing timeout T\textsubscript{\it{ageing}}}}\label{alg:MobCCN_A}
\end{algorithm}
%
%
\subsubsection{Multi-path mechanism}
\label{sub:MobCCN_R2}
\noindent
We introduce a protocol variant that adopts the same rule of MobCCN\_R1 to trigger a packet retransmission (i.e., the number of duplicated Interest packets goes above a defined threshold), but also leverages path diversity when retransmitting Interests. 
The idea here is that, instead of using the same path to retransmit Interests, the node explores a second path to find the content. To achieve this, the Interest to be retransmitted is sent to the second best forwarder.
The protocol works as follows (see Algorithm \ref{alg:MobCCN_R2_interest_processing}). As before, if multiple Interest packets for the same content arrive at one node, the node forwards only the first Interest to the node with the highest utility it knows (line 11-21), storing all the other ``faces'' in the corresponding PIT entry (line 6). 
However, if a retransmission event is triggered (line 7), the node retransmits the first Interest packet it has received to the second best forwarder, i.e., to the node with the second highest utility it knows (line 23-34). In this way, a second path is explored, hopefully increasing the probability of finding the content.
What we expect to obtain with this retransmission approach is an improvement of the delivery rate but at the cost of a higher end-to-end delay and hop count due to the presence of non-optimal paths.
In the following, we call this alternative solution ``\emph{MobCCN\_R2}''.
%
\subsection{Retransmission mechanisms based on timeout expiration}
\label{sub:retransmission-timeout}
\noindent
In the following sub-section we describe the two solutions based on periodic retransmission of the pending Interest packets. 
%

\subsubsection{Ageing mechanism}
\label{sub:MobCCN_A}
\noindent
The rationale behind this standard retransmission mechanism of the vanilla CCN that is adopted by MobCCN is to activate the retransmission mechanism according to the expiration of a specified timeout parameter. Basically, this version behaves in the same way of the MobCCN\_basic for what concerns the processing of Interest packets received by nodes as depicted in Figure \ref{fig:MobCCNprotocol:forwarding}, but the retransmission of pending Interests is executed periodically (see Algorithm \ref{alg:MobCCN_A}). 
Specifically, whenever the $ageing$ $timeout$ (T\textsubscript{\it{ageing}}) expires, each node $p$ checks its PIT table looking for still valid entries that can also be retransmitted (line 2). For those entries matching the above criteria, the retransmission of the Interest packet towards the node with highest utility occurs (line 3-14). Because of the timeout, the network has time to possibly reconfigure, and thus improve the accuracy of the ranking of forwarders towards the requested content. If there are old Interests, the corresponding entries are deleted from the PIT so as not become stale and be held indefinitely (line 16). What we want to assess here is the potential benefit of a periodic retransmission on protocol performance. Obviously, an increase of the traffic sent with respect to the MobCCN\_basic is expected, but this can be limited by tuning conveniently the ageing timeout. This solution is referred as ``\emph{MobCCN\_A}''.

\subsubsection{Ageing and hysteresis mechanisms}
\label{sub:MobCCN_AH}
\noindent
In this further version we use a combination of ageing and hysteresis mechanisms. Specifically, we enrich MobCCN\_A with a hysteresis mechanism aimed at improving the selection of the node to which to forward the Interest packets. Indeed, the encountered node is used as forwarded only if it improves by at least a threshold, called \emph{Hyst} value, the utility value of the node $p$ of forwarding the Interest to the newly encountered node. 
 Algorithms \ref{alg:MobCCN_AH} and \ref{alg:MobCCN_AH_timeoout} provide the description of the steps executed by node $p$ at the reception of an Interest packet and at the expiration of the ageing timeout, respectively. As it is apparent, the differences consist only in the condition that has to be satisfied for the transmission/retransmission of an Interest packet. Taking Algorithm \ref{alg:MobCCN_AH} as example, first the node $j$ with the highest utility value in FIB and CNU is identified. If $j$ is a neighbour of $p$, and the difference between the utility values of node $j$ and node $p$ is not less than a given threshold (determined by the hysteresis parameter), node $p$ forwards the Interest packet to the encountered node $j$ (lines 13-16), otherwise the Interest packet is stored waiting for the next contact (lines 17-18). 
 \setlength{\textfloatsep}{0pt}
\begin{algorithm}[t]
\scriptsize
		\begin{algorithmic}[1] 
			\Require $E_p$ \Comment{Set of current neighbours of node $p$}
			\Statex
			\State $i \gets Interest.\mathtt{Name}()$
			\If {$\mathrm{Data} \gets \mathrm{CS}.\mathtt{Find}(i)$}
				\State \Return Data 
			\EndIf
			\If {$\mathrm{PIT}.\mathtt{Find}(i)$}
				\State $\mathrm{PIT}.\mathtt{AddFace}(i,q)$
		    \Else
		    	\State $\mathrm{PIT}.\mathtt{CreateFace}(i,q)$
				\If {$\mathrm{FIB}.\mathtt{Find}(i)$}
					\State find the node $j$ with the highest utility for content type $i$ in FIB and CNU tables
					\If {$j \in E_p$ $\AND$ $U_{j,i} >  U_{p,i} * (1+ Hyst/100)$}
						\State forward Interest packet to node $j$
					\Else
						\State store Interest packet waiting for the next contact  
						
				\EndIf
				\Else
					\State drop Interest packet
				\EndIf
			\EndIf
		\end{algorithmic}
\caption{\small{MobCCN\_AH: processing of Interest packets received by node $p$}}\label{alg:MobCCN_AH}
\end{algorithm}
\setlength{\textfloatsep}{0pt}
\begin{algorithm}[t]
\scriptsize
		\begin{algorithmic}[1] 
				\Require $E_p$ \Comment{Set of current neighbours of node $p$}

		\Statex
			
		\ForEach{$e\in\mathit{PIT}$}
			\If {$ \NOT \mathrm{PIT}.\mathtt{Expired}(e)  $}
    				\State $i \gets \mathrm{PIT}.\mathtt{Name}(e)$
    				\If {$\mathrm{FIB}.\mathtt{Find}(i)$}
					\State find the node $j$ with the highest utility for content type $i$ in FIB and\\ \hspace{1.05cm} CNU tables
					\If {$j \in E_p$ $\AND$ $U_{j,i} > U_{p,i} * (1+ Hyst/100)$}
						\State forward Interest packet to node $j$
					\Else
						\State store Interest packet waiting for the next contact  
					\EndIf	

				\Else
					\State drop Interest packet
				\EndIf
			\Else
			    \State drop Interest packet
				\State PIT.remove(e)
			\EndIf
		
 		 \EndFor
			
		\end{algorithmic}
\caption{\small{MobCCN\_AH: procedure at node $p$ at the expiration of the ageing timeout T\textsubscript{\it{ageing}}}}\label{alg:MobCCN_AH_timeoout}
\end{algorithm}

The same condition holds at lines 10-18 of Algorithm \ref{alg:MobCCN_AH_timeoout}. The idea here is to verify if using a more conservative forwarding strategy improves the protocol performance. Postponing the Interest forwarding until a node whose utility value is significantly greater is found, is expected to guarantee a faster Interest convergence towards the content provider with a consequent decrease in paths length. In the rest of the paper this solution is referred to as ``\emph{MobCCN\_AH}''.
%

\section{Simulation setup}
\label{sec:simu-setup}
\noindent
To evaluate the performance of the proposed retransmission mechanisms, we extended the prototype of the MobCCN protocol described in \cite{Borgia:2016:MCP:2979683.2979695} with the enhancements described in Section \ref{sec:mobccn-enhancements}. We remind that MobCCN was implemented as an extension of CCN-lite~\cite{ccn-lite}, a lightweight open source CCN implementation supporting various hardware/software platforms, including the OMNeT++ simulator. 
\setlength{\textfloatsep}{5pt}
 \begin{table}[t]
 \footnotesize
   \centering
   \caption{Simulation settings for the basic scenario.}
   \label{tab:settings-basic}
   \begin{tabular}{l c}
     \toprule 

     \cellcolor[gray]{0.85}\textbf{Mobility parameters} & \cellcolor[gray]{0.85}\\
   
           \midrule
     area &   {1$km\times$1$km$ }\\
 
     mobile nodes & 60 \\

     mobile communities & 3 \\

     travellers & 3  \\

    communication range & 10$m$\\
    
    mobility model & {HCCM~\cite{Boldrini20101056} }\\

     avg. speed & {$\mathcal{U}(1,1.86)~m/s$} \\
     \toprule
          \cellcolor[gray]{0.85} {\textbf{Traffic parameters}}&  \cellcolor[gray]{0.85}\\
          \midrule
     producers & 6 \\
     
     consumers & 12, 24\\
     
     content types per producer & 4\\
     
     chunks per content type per producer & 5\\
     
     requests per consumers &40\\
     
     distribution req. times  & exponential ($\lambda=1/900$ ) \\
     
     simulation time  &  {36$~hours$ } \\
     
     req. start time  &  {11$~hours$ } \\
     
     req. end time  & { 22$~hours $} \\
     
     \bottomrule
      \end{tabular}
 \end{table}

\begin{figure}[t]
 	\centering
	\includegraphics[trim={0, 0, 0, 0}, clip, scale=0.35]{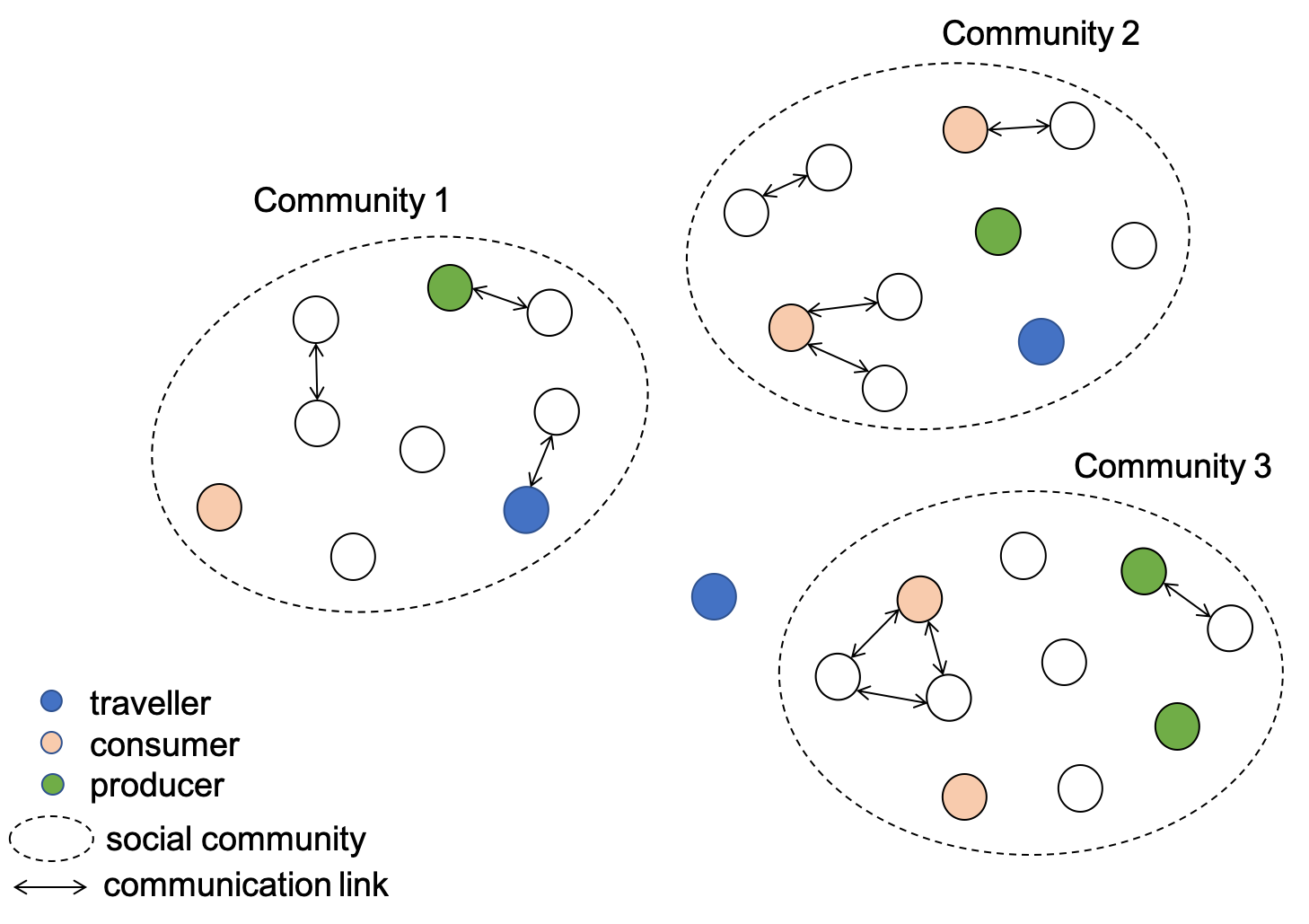}
	\vspace{-0.3cm}
 	\caption{Network overview.}

 	\label{fig:net-topo}
 \end{figure}

In the simulated environment, we investigate a basic scenario composed of 60 nodes randomly located in a square area of 1000mx1000m. We assume that a short-range wireless communication technology is used to opportunistically establish D2D links between nearby users' devices. This leads to a sparse network scenario with potentially short connection times, which is a challenging condition for any opportunistic data dissemination protocol (for details on simulation settings see Table~\ref{tab:settings-basic}). If not otherwise stated we assume that the users move with an average speed of 1.5m/s (a typical walking speed) according to the Home-cell Community-based Mobility Model (HCMM). HCMM is a popular human mobility model \cite{Boldrini20101056} able to reproduce realistic mobility patterns between the individuals of different social communities, as well as the opportunistic contacts of their personal devices. It is important to point out that we do not simulate the exact positions of nodes, but only the communication events between pairs of adjacent nodes using the starting and ending time instants of their opportunistic contacts. In addition, we do not simulate the communications between the static IoT devices and the users' mobile devices, but we consider as initial system state the time when mobile devices have already collected all the available contents from the IoT devices that are present in the area. In other words, we assume that when a requester issues an Interest packet for a specific content, that content is already stored in one of the available mobile devices. The direct utility of these nodes that initially store the content is set up to a randomly high value. Finally, we assume that a single copy of that content exists in the network when a request for that content is generated\footnote{This is equivalent to assume that the IoT devices cancel the content when it is delivered to one of the available users' devices.}. If not otherwise stated, in the following simulations mobile nodes are grouped into three separate communities, and each individual belongs to a single community and move within it. However, there is a small subset of nodes, called ``travellers'', which can sporadically visit external communities and allow the dissemination of contents between communities. Without loss of generality, in our study we assume that each community has a single traveller node. This setup represents, for example, users moving in a specific (social) environment, such as their working place, and sometimes moving in other contexts, as a side effect of ``external" social relationships. A schematic representation of the resulting network is shown in Figure \ref{fig:net-topo}.

Regarding the traffic patterns, each content producer manages 4 content types, each composed of 5 different chunks, thus storing 20 contents in total, while each content consumer asks for 40 different contents. Among these 40 requests, half of them targets contents that are produced by nodes belonging to the same community of the requester, while the remaining 20 requests are uniformly distributed among the contents produced by the other two external communities. Content requests are generated after a warm-up period and terminate before the simulation end to reduce the probability that packet losses occur because the simulation ends before the data dissemination process is concluded. Content requests are generated following an exponential distribution of inter-request times with parameter $\lambda=1/900$, which corresponds to a content request every 15 minutes, on average.    

As anticipated, in Section \ref{sec:perf-eva} we compare all the MobCCN retransmission mechanisms with benchmarks under different scenarios and working hypothesis. It is useful to remark that the low speed and the size of the simulated area yields to long inter-contact times, which leads to long data dissemination delays. However, delays will scale down is denser network scenarios and/or higher mobility speeds would be considered. 

%

%
\subsection{Performance metrics}
\label{sub:performance-metrics}
 \noindent
The performance comparison is based on the following network-related performance metrics: 
\begin{itemize}[leftmargin=*,itemsep=0pt,parsep=0pt]
    \item \emph{Delivery rate}: percentage of Data packets successfully received at the consumers. In scenarios where we assume an infinite bandwidth during contacts among nodes\footnote{This is customary in the opportunistic networking literature.}, the only cause of unsuccessful Data delivery is ascribed to the failure of MobCCN in delivering Data packets to consumers before the end of the simulation;
    \item \emph{End-to-end latency}: time interval between the generation of the Interest packets and the successful reception of the corresponding Data packets (this metric is computed only for those Data packets which have been correctly received by consumers);
    \item \emph{Hop count}: number of nodes through which Data packets pass in their path from producers to consumers\footnote{As Data packets follow the reverse path of their corresponding Interest packets, the hop count also represents the number of hops of Interest packets.}; 
    \item \emph{Total traffic}: the total amount of traffic generated by all the nodes. This includes the forwarding traffic composed by Interest and Data packets, and the control traffic due to routing when present. 
\end{itemize}
To better highlight the impact of mobility on system performance, we also measure delivery rate, end-to-end delay and hop count at the ``community level''. Specifically, for the aforementioned metrics we measure two different components: i) a ``local" one, which refers to the metric result achieved for the requests of contents that are produced by nodes within the same community of the requester, and ii) an ``external" one, which is related to the metric result obtained for the requests of contents held by nodes outside the requester's community.     

To quantify the impact of retransmission traffic we introduce the following metrics: 
\begin{itemize}[leftmargin=*,itemsep=0pt,parsep=0pt]
\item \emph{Retransmission per Interest}: number of retransmitted Interests per generated requests, which measures the average amount of retransmissions ($I_{reTX}$) for each Interest generated by a consumer ($I_{gen}$). It is given by:
\begin{equation}
	RPI =  \frac{I_{reTX}}{I_{gen}} \; .
	\label{eq:RI}
\end{equation}
%
\item \emph{Retransmission per Data}: number of retransmitted Interests per received content, which measures the average amount of retransmissions needed to successfully receive a Data packet at the consumer. It is given by:
\begin{equation}
	RPD =  \frac{I_{reTX}}{D_{rcv}} \; .
	\label{eq:RD}
\end{equation}
\end{itemize}
Note that $D_{rcv}$ can be lower than $I_{gen}$ due to a delivery rate lower than one.

Each simulation is replicated 10 times and the performance results are averaged over all the replicas. In the following, average values are shown with $95\%$ confidence interval. 
%
\subsection{Benchmarking protocols}
\label{sub:benchmarking-protocols}
\noindent
We compare MobCCN and its variants with two benchmarks based on the well-known Epidemic routing protocol for opportunistic networks~\cite{Vahdat00epidemicrouting}. The first benchmark is the classical Epidemic scheme that works by flooding both Interest and Data packets in the network, i.e., each time two nodes encounter they exchange all their Interest and Data packets. Intuitively, under the simplifying hypothesis of infinite bandwidth and unlimited cache size at the nodes, this protocol provides the shortest and minimum-delay path from a source to a destination, representing an upper bound in terms of delivery rate, and a lower bound in terms of end-to-end latency and hop count. Its major drawback is the high cost in terms of network overhead due to the uncontrolled packet flooding. Including Epidemic in the comparison allows us to understand how much our proposed solutions deviate from the optimal one in terms of delivery rate, end-to-end latency and hop count, and how much they gain in terms of network traffic. In the rest of the paper we refer to it as ``\emph{Ideal Epidemic}".  

The second benchmark is a \emph{copy-limited} version of Epidemic, which works as follows. Whenever an intermediate node receives an Interest packet, it checks if the Interest packet has been already forwarded. If this is the case, it simply drops the packet, otherwise it forwards it with a probability \emph{r} (in our experiment we set \emph{r} to 0.5). Data packets are then routed back to those nodes who requested them exploiting the reverse-paths, as happens for MobCCN. As apparent, this version of Epidemic protocol drastically limits the number of Interest and Data packets that are flooded in the network at the cost of a reduced delivery rate and an increased end-to-end latency and hop count. Including this protocol in the comparison allows to investigate which are pros and cons of the utility-based protocols compared to the flooding-based ones. In the rest of the paper we refer to it as ``\emph{Limited Epidemic}".
%
%
\section{Performance analysis} \label{sec:perf-eva}
\noindent
In this section we compare the performance of the different retransmission mechanisms described in Section \ref{sec:mobccn-enhancements} against the two benchmark protocols described in Section \ref{sub:benchmarking-protocols}, namely Ideal Epidemic and Limited Epidemic. 

 \begin{figure}[t]
 	\centering
	\includegraphics[angle=-90,width=0.35\textwidth]{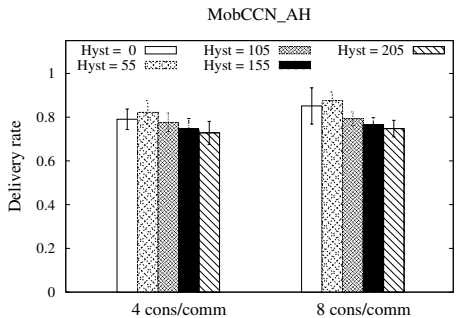}
	\vspace{-0.2cm}
 	\caption{Sensitivity of MobCCN\_AH on ${Hyst}$: delivery rate in 4 consumers/community and 8 consumers/community configurations ($T_{ageing}=40000$).}
 	\label{fig:performance_sens_AH:delRate_AH}
 \end{figure}
 \begin{figure}[t]
 	\centering
	\includegraphics[angle=-90,width=0.38\textwidth]{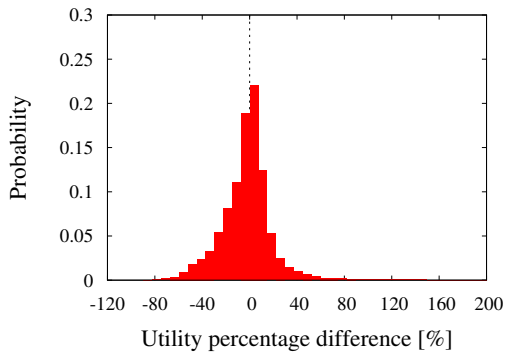}
	\vspace{-0.2cm}
 	\caption{Distribution of the utility percentage differences among nodes. }
 	\label{fig:utility-percentage-difference}
 \end{figure}
Before that, we briefly summarise the sensitivity analysis we conducted to fine-tune the protocol parameters. The interest readers may refer to \cite{Borgia:2020} for the complete analysis.   
In the retransmission mechanisms based on the number of received requests, namely MobCCN\_R1 and MobCCN\_R2, the parameter that triggers the retransmission of duplicate Interest packets is the retransmission threshold $ReTX_{threshold}$. Our results show that different $ReTX_{threshold}$ values have a small impact on the system performance. We have also noted that there is a small increase of the delivery rate when each duplicate reception of an Interest packet triggers a retransmission, hereafter referred to ``0'' setting, while the other performance metrics remain almost unchanged. 
As far as MobCCN\_A is concerned, we observed significant difference for the delivery rate (up to 10 percentage points) when varying the parameter that regulates the periodic retransmission of pending Interest packets. It is clear that a sub-optimal setting of the ageing parameter results into a worse protocol performance. In our scenario the best configuration is $T_{ageing}=40000$. Concerning the parameters specific to MobCCN\_AH, i.e., the hysteresis percentage ${Hyst}$, we observed that the delivery rate initially increases, reaches a peak value and, then, start decreasing (see Figure \ref{fig:performance_sens_AH:delRate_AH}). The maximum delivery rate is achieved with a hysteresis percentage equal to 55\%. This means that the strategy of delaying the retransmission of the pending Interests is beneficial, but up to a certain extent. The delivery rate gain is up to 5\% in such cases. On the contrary, postponing too much the forwarding in order to find a node whose utility is significantly higher is counterproductive. For instance, waiting for a node whose utility is two times higher leads to lower delivery rates, even lower than what is obtained with MobCCN\_A (first column). This can be explained by observing the distribution of utility percentage differences between encountered nodes in Figure \ref{fig:utility-percentage-difference}, where it is shown that the majority of nodes have utility at most one time higher, while most of the distribution lies within the range [- 50\%, 30\%]. Thus, setting the hysteresis threshold to a high value may have the negative effect of introducing an excessive delay before triggering a retransmission of an Interest packet.

\begin{table}[h!]
 \footnotesize
   \centering
   \caption{MobCCNs settings.}
   \label{tab:MobCCNsettings}
   \begin{tabular}{l c}
     \toprule 
     \cellcolor[gray]{0.85}\textbf{MobCCN version} & \cellcolor[gray]{0.85}\textbf{parameter}\\
      \midrule
     MobCCN\_R1 &  ReTX\textsubscript{\it{threshold}} = 0\ \\
     MobCCN\_R2 & ReTX\textsubscript{\it{threshold}} =  0\ \\
     MobCCN\_A  & T\textsubscript{\it{ageing}} = 40000\ \\
     MobCCN\_AH & T\textsubscript{\it{ageing}} = 40000, Hyst  =  55\%\ \\
         \bottomrule
      \end{tabular}
\end{table}
%

In the rest of the section we compare the performance of all the MobCCN variants against the two benchmarks taking into account the above results. Specifically, each MobCCN variant is simulated using the best configuration for its protocol parameter(s) that is obtained when the delivery rate is maximised. Table \ref{tab:MobCCNsettings} summarises the setting used.

%
 \begin{figure*}[t]
 	 \hspace{-6mm}
 	 \mbox{
	\includegraphics[trim={30cm 0cm 3cm 0.5cm}, angle=-90,width=1.03\textwidth]{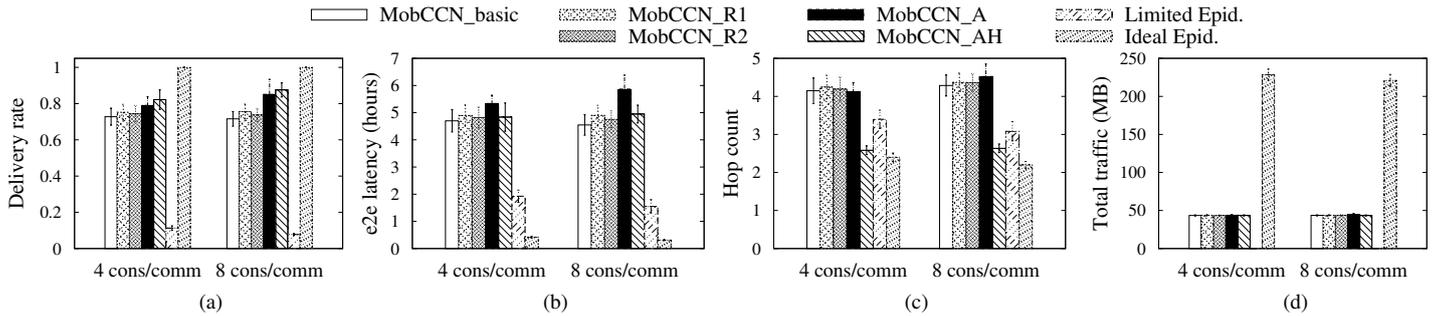}
	}
	\vspace{-0.3cm}
 	\caption{Network-level performance: delivery rate (\ref{fig:netw_performance}a), end-to-end latency (\ref{fig:netw_performance}b), number of hops (\ref{fig:netw_performance}c), and total traffic (\ref{fig:netw_performance}d) in 4 consumers/community and 8 consumers/community configurations.}

 	\label{fig:netw_performance}
	\vspace{-0.3cm}
 \end{figure*}

We split the evaluation into five parts. In Section \ref{sub:basic-scenario} we focus on the basic scenario, whose parameter settings are depicted in Table \ref{tab:settings-basic}. Here, we first focus on the aggregate behaviours of the protocols by analyzing the metrics at the network-level. Then, we increase the detail of the analysis by investigating what happens at the community level. To investigate the protocols performance under different traffic patterns, in Section \ref{sub:load-increase-scenario} we change the number of consumers requesting content (i.e., increase of traffic load), while in Section \ref{sub:different-request-pattern-scenario} we analyse a scenario where there is an overlapping between content requests of different consumers (i.e., increase of content reuse). In Section \ref{sub:retransmission-performance} we show more detailed results on the amount and distribution of retransmission traffic generated by the MobCCN retransmission mechanisms. 

It is important to point out that the results in Sections \ref{sub:basic-scenario}-\ref{sub:retransmission-performance} are obtained under the simplified hypothesis of infinite communication bandwidth, as we wanted to measure the number of failed content requests due to protocol inefficiency and mobility patterns without the effect of bandwidth bottlenecks. In Section \ref{sub:lim-bw-performance} we relax such assumption and we also investigate the system performance under limited network bandwidth.

%
\subsection{Basic scenario}
\label{sub:basic-scenario}
\vspace{0.3cm}
%
\subsubsection{Network-level performance}
\label{sub:network-level-performance-scenario3}
\noindent
Figures \ref{fig:netw_performance} illustrates the obtained results for all the performance metrics at the network level in the basic scenario. Looking at the delivery rate (see Figure \ref{fig:netw_performance}a), Ideal Epidemic achieves the best performance with a delivery rate equal to 100\%, as expected. By contrast, Limited Epidemic results to be the worst scheme with a delivery rate below 10\%. The main reason for such a low performance is the probabilistic strategy used for forwarding Interest packets, which can lead to useless transmissions of the same Interest packet. Conversely, in the same conditions, MobCCN significantly outperforms Limited Epidemic with delivery rates higher than 70\%, thus confirming that the use of the utilities provides a more efficient way of searching for requested content. 
As far as the MobCCN variants, we observe the following delivery rate for the ``4 cons/comm" configuration: 75\% for MobCCN\_R1, 74\% for MobCCN\_R2, 79\% for MobCCN\_A, and 82\% for MobCCN\_AH.
The figure clearly highlights that transmission reliability greatly improves with retransmissions, resulting in an increase of delivery rates. However, such advantages are marginal if the retransmission strategy relies only on the number of the received requests. This is independent of the type of paths followed, either be they optimal as in MobCCN\_R1 or sub-optimal as in MobCCN\_R2. 
Indeed in such cases, since the retransmissions follow the packet arrival rate, they occur close together in time (within two hours from each other). As consequence, the utility values of the possible forwarders that can be used to find a requested content change little. This means that, in case of MobCCN\_R1, the forwarder with the highest utility towards a specific content is likely to remain the same node of the first transmission. As far as MobCCN\_R2, a different node is picked at least at the first hop, as the second highest utility is used. But, in general, the paths used by retransmitted packets are mostly overlaid with those used in the first transmission. 
On the contrary, advantages are more significant if the retransmission is executed on a periodic basis, as the delivery gains are in the range of 10\%-15\%. This is due to the fact that, in this second case, as retransmissions occur after longer times, the network reorganizes itself and the nodes identify new best forwarders. Therefore, retransmissions explore new paths, separated from those used in the first transmission. 
It is worth pointing out that, among the two periodic-based protocols, MobCCN\_AH reaches better performance. The hysteresis mechanism indeed ensures that the nodes look for the nodes to which to forward/retransmit their requests more carefully, forwarding them only when there is a significant advantage in terms of utility.
Another observation is related to the increase of the delivery rate when the number of consumers grows from four to eight. The more the consumers, and the more the number of content requests to satisfy, but the higher is also the probability that requests are partially overlapped. Therefore, one retransmission may satisfy more requests at the same time due to the aggregation mechanism of Interest packets used in the PIT. Again, MobCCN\_AH outperforms the other MobCCN versions.

Average e2e latencies are depicted in Figure \ref{fig:netw_performance}b. Also in this case, Ideal Epidemic outperforms all the other protocols experiencing very low delays. This is expected as, in order to find the requested contents, it spreads the Interests on all the nodes encountered. Low delays are also experienced by Limited Epidemic, but this depends mainly on two factors: i) this average is computed from a low number of successfully received Data packets, and ii) those Data packets mainly refer to contents belonging to the local consumers' communities and thus experience shorter latencies. This aspect will be further examined in the next section. As far as MobCCN and its variants, e2e latencies are in the range of $[4.7-5.3]$ hours in the ``4 cons/comm" configuration and in the range of $[4.55-5.86]$ hours in the ``8 cons/comm" configuration. Concerning the two periodic-based protocols, we observe  that MobCCN\_A has the largest average latency to retrieve the contents, while MobCCN\_AH performs better with a latency only a bit higher than MobCCN\_basic. 
Also in this case, we observe that the hysteresis mechanism helps in improving the performance lowering the end-to-end latencies.

In Figure \ref{fig:netw_performance}c the average hop count is reported. As far as this metric, the graph highlights an important MobCCN\_AH feature. On the one hand, it outperforms all the other MobCCN versions by cutting the paths length by half. On the other hand, its performance is almost equivalent to that of Ideal Epidemic, with about 2.5 hops on average. This is an important result that highlights the ability of MobCCN\_AH to achieve excellent delivery rates with low resource consumption, in contrast to Ideal Epidemic that, on the contrary, needs to replicate packets over the network, hence wasting useful resources.

Figure \ref{fig:netw_performance}d shows the total traffic for all the protocols. We recall here that the total traffic metric includes all the traffic generated by all the nodes, thus for the MobCCN versions this includes not only the Interest and Data packets as in Ideal Epidemic or Limited Epidemic, but also the control traffic due to the routing protocol. As shown by the figure, Ideal Epidemic has the largest total traffic which can be as high as 230MB, followed by all the MobCCN versions with less than 50MB, and by the Limited Epidemic, whose traffic is almost negligible.  The MobCCN versions therefore perform very well, significantly reducing network traffic compared to Ideal Epidemic, with a 78\% traffic cut. Obviously, they produce more traffic over the network than Limited Epidemic, mainly due to the routing information that needs to be be sent in order to populate FIB tables, but it is a fair compromise that permits to obtain good delivery rates (see Figure \ref{fig:netw_performance}a). 

 \begin{figure}[t]
 \hspace{-1cm}
 	\centering
	\includegraphics[trim={0.5cm 0cm 1cm 2cm},width=0.45\textwidth]{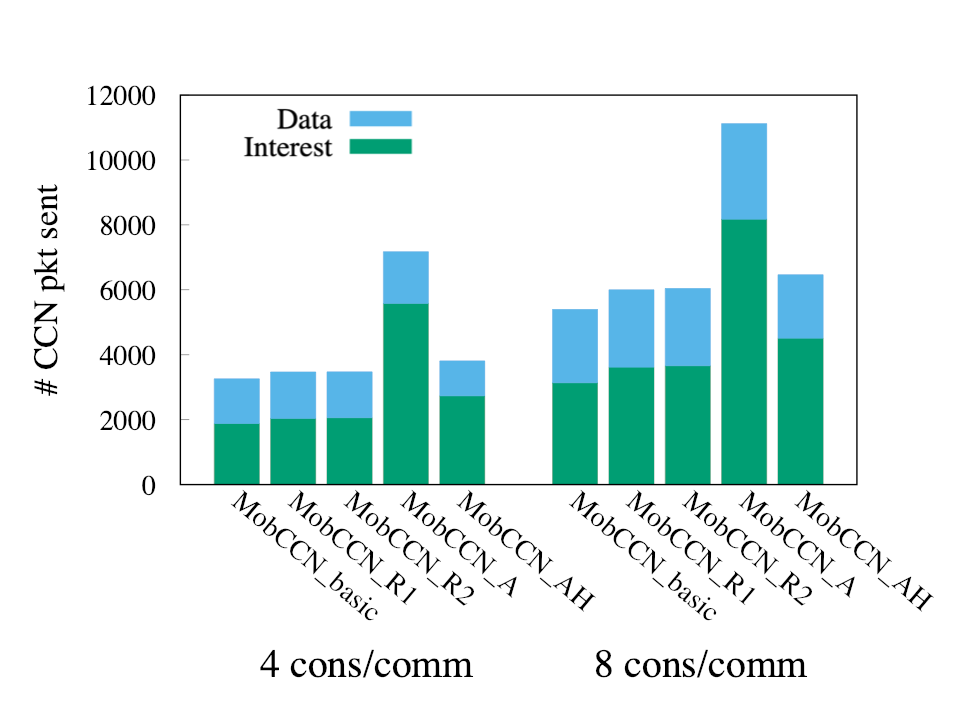}
	\vspace{-0.5cm}
 	\caption{Breakdown of the forwarding component for the MobCCN protocols.}

 	\label{fig:netw_performance:aggr_forw_pkts}
 \end{figure}

Finally, to investigate more in depth the impact of the retransmission mechanisms on the protocol overheads, Figure \ref{fig:netw_performance:aggr_forw_pkts} shows the number of Interest packets and Data packets that are produced by each MobCCN variant. We note that MobCCN\_basic sends the lowest number of Interests, followed by MobCCN\_R1 and MobCCN\_R2 with similar aggregate number of forwarding traffic, and then by MobCCN\_AH with a 18\% increase compared to MobCCN\_basic. On the contrary, MobCCN\_A has the worst performance by doubling the number of Interests with respect to the other MobCCN versions. On the contrary, the behaviour of Data packets is different. In this case, MobCCN\_A still shows the worst performance but the difference with the other protocols is less remarkable. MobCCN\_AH achieves the better performance reducing the number of Data sent by 17\% with respect to MobCCN. Reducing the number of Data packets as done by MobCCN\_AH is very useful, and such advantage grows as the data transmitted increases. This essentially happens for two reasons. First, since the size of a Data packet is on average much greater than the one of Interest packets, as it consists of a few bytes mainly related to the identification string of the content to search, the Data component plays a key role compared to the Interest one. Thus, a reduction of the number of transmitted Data packets corresponds to a significant decrease in the amount of forwarding traffic sent compared to other protocols. Second, as soon as the data traffic intensifies, the forwarding component becomes predominant over the routing one. There is therefore a consequent reduction of the overhead necessary to guarantee the correct population of the FIB tables, and more generally of the total traffic sent.
 \begin{figure}[t]
 	\centering
 	\vspace{-0.5cm}
	\includegraphics[trim={2cm 0.5cm 0cm 0.5cm}, angle=-90,width=0.52\textwidth]{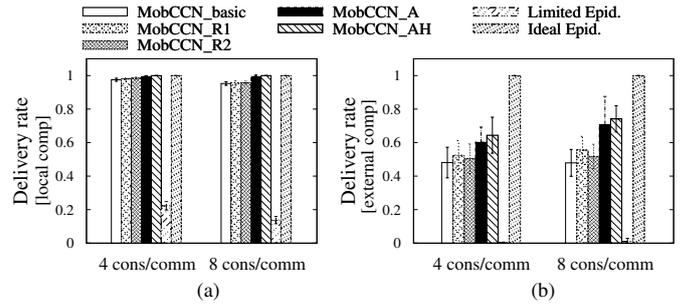}
	\vspace{-0.3cm}
 	\caption{Community-level performance: delivery rate for the local community (\ref{fig:deliveryRate_LocalForReq_scen3}a), and the external communities (\ref{fig:deliveryRate_LocalForReq_scen3}b)}

 	\label{fig:deliveryRate_LocalForReq_scen3}
 \end{figure}
%

%
\subsubsection{Community-level performance}
\label{sub:community-level-performance-scenario3}
\noindent
In this section we focus on the protocols performance at the community level, that is, as described in Section \ref{sub:performance-metrics}, we split the performance indexes in two components according to the membership of the requested contents, be they local or external, and we analyse them separately. We recall here that every consumer makes 40 content requests such that 20 are related to contents inside its own community and the remaining 20 are uniformly distributed among the other two external communities. 

 \begin{figure}[t]
 	\centering
	\includegraphics[angle=-90,width=0.4\textwidth]{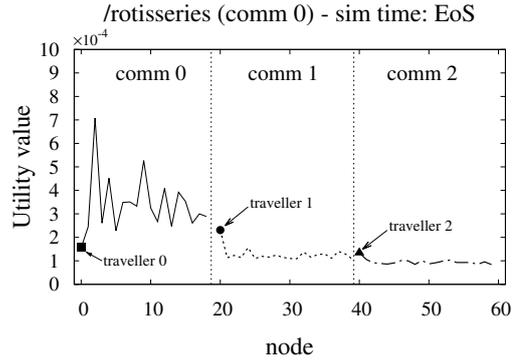}
	\vspace{-0.3cm}
 	\caption{Example of the utility values towards a content belonging to community 0 (i.e., ``rotisseries")  at the end of the simulation. Nodes of community 0, which are shown on the left side of the figure, experience high non-uniform utility values as they meet directly the content producer but with different mobility patterns. The utility value of the traveller of community 0 is, on the contrary, the lowest because it also visits communities 1 and 2, thus reducing the probability to meet the content producer. Similar utilities are experienced also by travellers of communities 1 and 2, which commute by definition among all the three communities. Nodes outside community 0 experience very low utilities because they never meet directly the content provider. Note that all the three travellers have utility values of about 50\% higher than the utilities of nodes within community 1 and 2.}

 	\label{fig:utility}
 \end{figure}
%

Figure \ref{fig:deliveryRate_LocalForReq_scen3} shows the two separated delivery rate components: the local one in Figure \ref{fig:deliveryRate_LocalForReq_scen3}a, and the external one in Figure \ref{fig:deliveryRate_LocalForReq_scen3}b. As highlighted by the figure on the left, Limited Epidemic has the lowest performance, with delivery rates of 22\% and 13\% in the ``4 cons/comm" and ``8 cons/comm" configuration, respectively. This is due to the fact that in Limited Epidemic nodes are all equivalent and the search of the forwarder to which to send the request is random. On the contrary, the delivery rate performance rises significantly when MobCCN and its variants are used because here nodes have different utility weights. MobCCN alone (i.e., MobCCN\_basic) reaches values above 95\% in both configurations, confirming that the utility mechanism is a successful strategy. Also, when retransmission mechanisms are running, the protocol reliability further increases, with \\MobCCN\_AH that reaches a 100\% delivery rate retrieving all the contents belonging to the local community of consumers, equal to Ideal Epidemic.
Similar considerations can be drawn for the external component of the delivery rate as depicted by Figure \ref{fig:deliveryRate_LocalForReq_scen3}b, although the numbers involved are lower. The unique value that remains unchanged is the one related to Ideal Epidemic, which is able to deliver successfully all the contents requested by nodes in external communities. By contrast, Limited Epidemic's performance drops further with only one satisfied request, on average. 
Concerning the other protocols, we observe that at least 50\% of the external content requests are satisfied with MobCCN and its variants. In addition, this figure clearly highlights that the retransmission mechanisms introduced, although useful for performance, affect differently the delivery rate. Specifically, compared to MobCCN\_basic, MobCCN\_R1 and MobCCN\_R2 increase the delivery rate by a few percentage points only. On the contrary, the delivery gain obtained with MobCCN\_A and MobCCN\_AH is higher, in the order of [12-30] percentage points. Further observations can be then derived for each retransmission scheme. First, MobCCN\_R1 performs slightly better than MobCCN\_R2, that is, retransmitting over the same best path guarantees higher performance with respect to exploiting multi-paths that pass through non-optimal forwarders, as happen in MobCCN\_R2. Second, MobCCN\_AH has the best performance with delivery rates equal to 64\% and 74\% in the ``4 cons/comm" and ``8 cons/comm" configuration, respectively. This result indicates that combining a periodic retransmission based on a large retransmission timeout and a hysteresis mechanism significantly increases the transmissions reliability, especially for those contents belonging to external communities. 
In this scenario, indeed, it is essential to detect those nodes that can contribute the most to the content dissemination process, that is, the \emph{travellers}, which are the only mobile devices able to retrieve external contents as they commute among communities. As shown in Figure \ref{fig:utility}, the travellers have utility values towards contents of external communities much higher than the corresponding utilities of nodes inside the local community (of about 50\%). By measuring the utility difference among the nodes in contact, as performed by the hysteresis condition, MobCCN\_AH is able to establish whether or not the encountered node is a traveller. When the utility difference is minimal, the node in contact is a common node. On the contrary, when this difference becomes significant and higher than the hysteresis value, the node in contact is a traveller. The latter is then chosen by the forwarding process and exploited to retrieve the desired external contents.

 \begin{figure}[t]
 	\centering
	\includegraphics[trim={2cm 0.5cm 0cm 2cm}, angle=-90,width=0.52\textwidth]{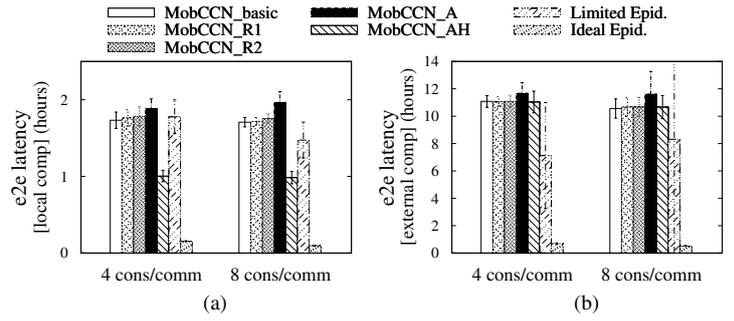}
	\vspace{-0.3cm}
 	\caption{Community-level performance: end-to-end delay for the local community (\ref{fig:e2e_LocalForReq_scen3}a), and the external communities (\ref{fig:e2e_LocalForReq_scen3}b)}

 	\label{fig:e2e_LocalForReq_scen3}
 \end{figure}
 \begin{figure}[t]
 	\centering
	\includegraphics[trim={2cm 0.5cm 0cm 0.5cm}, angle=-90,width=0.52\textwidth]{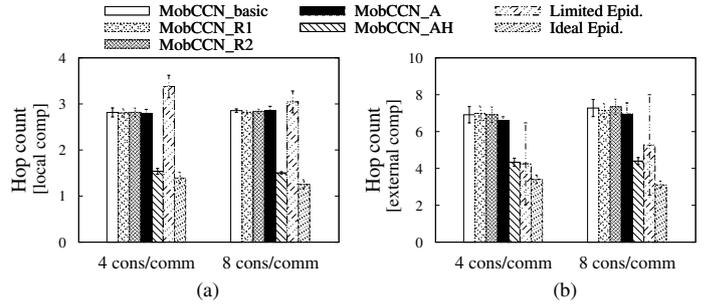}
	\vspace{-0.3cm}
 	\caption{Community-level performance: hop count for the local community (\ref{fig:nHop_LocalForReq_scen3}a), and the external communities (\ref{fig:nHop_LocalForReq_scen3}b)}

 	\label{fig:nHop_LocalForReq_scen3}
 \end{figure}
 \begin{figure*}[t]
 	 \hspace{-6mm}
 	 \mbox{
	\includegraphics[trim={30cm 0cm 3cm 0.5cm}, angle=-90,width=1.03\textwidth]{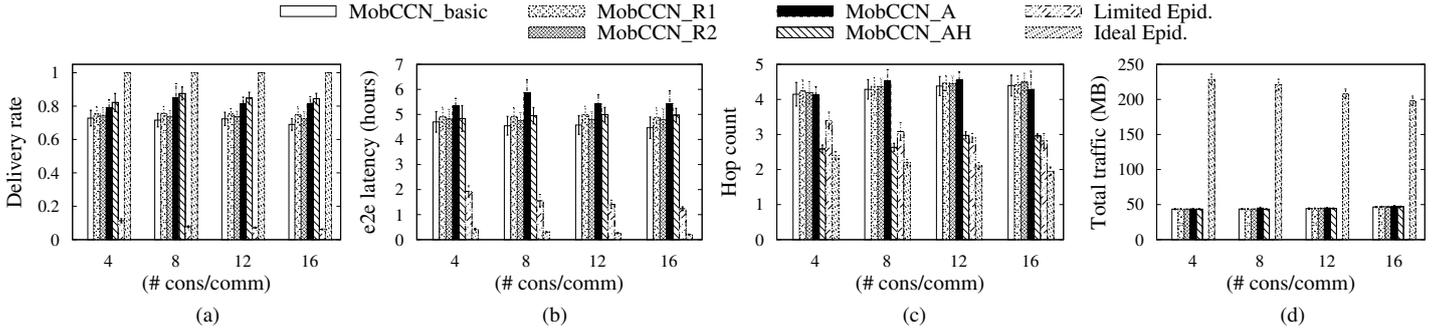}
	}
	\vspace{-0.3cm}
 	\caption{Network-level performance in the scenario with an increased network load: delivery rate (\ref{fig:netw_performance-load}a), end-to-end latency (\ref{fig:netw_performance-load}b), number of hops (\ref{fig:netw_performance-load}c), and total traffic (\ref{fig:netw_performance-load}d) for different consumers/community configurations.}

 	\label{fig:netw_performance-load}
	\vspace{-0.2cm}
 \end{figure*}

Figure \ref{fig:e2e_LocalForReq_scen3} shows the two end-to-end latency components. For what concerns the local component, Ideal Epidemic introduces the smallest average time to satisfy the data requests, followed by MobCCN\_AH with one hour on average, and then by all the other protocols. When the number of Interest requests increases, the network load becomes higher, however the latency delays show no significant change. Note also that MobCCN\_AH halves the latency with respect to MobCCN\_A, highlighting once again its advantages. For what concerns the external components, Ideal Epidemic, by replicating Interests on every encountered node, experiences very low latencies also to retrieve external contents. On the contrary, all the MobCCN variants have similar performance but experiencing higher delays. Also in this case, MobCCN\_AH slightly lowers the delays with respect to MobCCN\_A. Limited Epidemic latency is statistically insignificant as it refers to a single sample (the external delivery rate is near 0, as shown by Figure \ref{fig:deliveryRate_LocalForReq_scen3}b).   

Finally, in Figure \ref{fig:nHop_LocalForReq_scen3} the two components of the hop count metric are depicted.       
Both components show a similar behaviour, both when retrieving local contents and external contents. Obviously, the paths used to retrieve local contents are shorter than those for retrieving external contents. Overall, the group of protocols with the shortest paths length is composed by MobCCN\_AH and Ideal Epidemic, followed by the group composed of the remaining MobCCN protocols (i.e., MobCCN\_basic, MobCCN\_R1, MobCCN\_R2 and MobCCN\_A) that achieve similar performance. Finally, Limited Epidemic has the worst performance using the longest paths (note that also here the foreign component is insignificant as it refers to a single sample). 
This figure highlights the ability of MobCCN\_AH to halve the paths length compared to all the other MobCCN solutions, getting closer to the minimum paths detected by Ideal Epidemic, all this by lowering significantly the overall network traffic. 
Therefore, postponing the transmission in order to find the optimal forwarder avoids the unpleasant ping-pong situations among nodes, and allows to exactly identify those nodes useful for a successful Data recovery, such as for example the travellers for retrieving external contents. As a result, paths are shorter and resources consumption is reduced.

Also this community-level analysis confirms that a smart retransmission mechanism helps in improving the MobCCN performance. Specifically, we observe that MobCCN\_AH, with its combination of periodic-based retransmission and hysteresis mechanism, is able to achieve good performance results getting very close to the Ideal Epidemic performance for satisfying requests of local contents. Moreover, it achieves the best performance with respect to the remaining MobCCN variants when retrieving contents of external communities. This is also a good result given that the contacts between the communities, consisting of the travellers movements only, are very limited since there is only 1 traveller per community.

%
\subsection{Impact of network load}
\label{sub:load-increase-scenario}
\noindent
Figure \ref{fig:netw_performance-load} illustrates the performance of the considered protocols at the network level when varying the network load. Specifically, we increase the number of consumers generating Interests from 4 up to 16 consumers per community. 
 \begin{figure*}[t]
 	 \hspace{-6mm}
 	 \mbox{
	\includegraphics[trim={30cm 0cm 3cm 0.5cm}, angle=-90,width=1.03\textwidth]{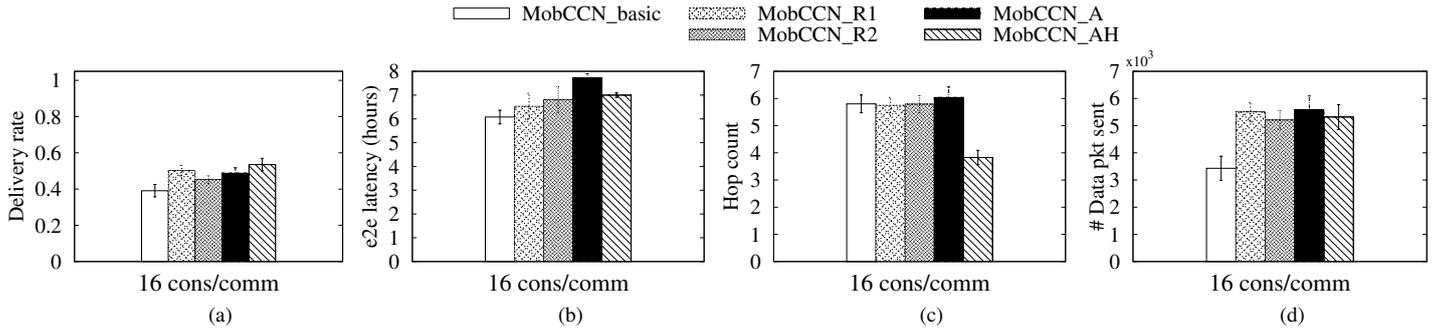}
	}
	\vspace{-0.3cm}
 	\caption{Network-level performance in the scenario with a different request pattern: delivery rate (\ref{fig:netw_performance-req-pattern}a), end-to-end latency (\ref{fig:netw_performance-req-pattern}b), number of hops (\ref{fig:netw_performance-req-pattern}c), and total traffic (\ref{fig:netw_performance-req-pattern}d) in the 16 consumers/community configuration.}

 	\label{fig:netw_performance-req-pattern}
	\vspace{-0.6cm}
 \end{figure*}

Looking at Figure \ref{fig:netw_performance-load}a, we observe that in general the periodic retransmission approach continues to perform better than the retransmission approach based on the number of received requests also when the traffic load increases, with an average gain in the range of [10-15]\%. Again, MobCCN\_AH achieves higher performance than the MobCCN\_A counterpart, hence confirming its efficiency in the different network conditions we tested. Figure \ref{fig:netw_performance-load}b shows that MobCCN\_A experiences higher end-to-end latencies with respect to the rest of MobCCN variants, which are, on the contrary, almost aligned around 5$h$. MobCCN\_basic latencies are lower but this is essentially due to a lower number of received contents. As far as the hop count metric (see Figure \ref{fig:netw_performance-load}c), we observe that MobCCN\_AH significantly reduces the path lengths with respect to the other MobCCN retransmission schemes, with a reduction of around 33\%. In Figure \ref{fig:netw_performance-load}d we observe that the total traffic for all the MobCCN versions is basically the same and stabilized around less than 50MB. When increasing the number of consumers, we also note a slight rise of the traffic, which is basically due to the higher number of Interests generated. On the contrary, Ideal Epidemic exhibits a decreasing trend. The same behaviour is apparent also for the other performance metrics. Since there are overlapping requests, they spread even faster with the growing number of consumers. This is a consequence of how Interests are replicated at nodes. Indeed, in Ideal Epidemic, a node floods Interests to all the nodes it is in contact with. Therefore, the greater is the number of the requests for the same content, the higher is the speed with which that Interest is spread in the network.  As a result, in Ideal Epidemic contents are retrieved faster, cross fewer hops, and the network traffic is slightly reduced.

As far as the performance at the community level, we observe a similar trend under these scenarios, and thus we omit the corresponding plots here.
%
\subsection{Impact of content request pattern}
\label{sub:different-request-pattern-scenario}
\noindent
In this section we report the performance results of the scenario in which we vary both the pattern of content requests and the traffic load. Specifically, we analyse a scenario with 16 consumers per community where each consumer asks for the same set of contents, that is, content requests are completely overlapped. The requested set consists of 80 different contents divided, as before, between their own community (40) and the two external ones (40).  

Looking at the delivery rates shown in Figure \ref{fig:netw_performance-req-pattern}a, we note that they are significantly lower than those in the previously considered scenarios being in the range [0.4, 0.55]. However, \\MobCCN\_AH continues to achieve the highest delivery performance with a delivery gain of 15\% with respect to MobCCN\_basic. As far as the e2e latencies (see Figure \ref{fig:netw_performance-req-pattern}b), they range in the [6$h$-8$h$] interval. MobCCN\_A experiences again the highest delays for retrieving content, followed by MobCCN\_AH  and MobCCN\_R2 with similar values, then MobCCN\_R1 and finally  MobCCN\_basic. However, MobCCN\_basic experiences also the lowest number of received contents. Figure \ref{fig:netw_performance-req-pattern}c, which shows the average path length, highlights once again the capability of MobCCN\_AH to reduce it by more than 30\% with respect to the other protocols. Finally, as a measure of the network traffic, in Figure \ref{fig:netw_performance-req-pattern}d we report the forwarding traffic component related to the Data packets sent, being the one that influences the most the amount of forwarding traffic sent. Excluding MobCCN\_basic, whose lower traffic is mainly due to the lower delivery rate, also in this case MobCCN\_AH achieves good performance reducing the number of Data packets sent with respect to the other protocols. 

This scenario, which is in general advantageous for protocols that make use of in-networking caching (even temporarily), confirms the ability of MobCCN\_AH to find the desired contents more precisely and with higher reliability than the others.

\begin{figure}[tbp] 
\centering
\vspace{-0.4cm}
\hspace{-0.75cm}
    \subfloat[\label{fig:retransmission-metrics:RPI}]{%
      \includegraphics[trim={0cm 0.5cm 0cm 0.6cm},clip,angle=-90,width=0.255\textwidth]{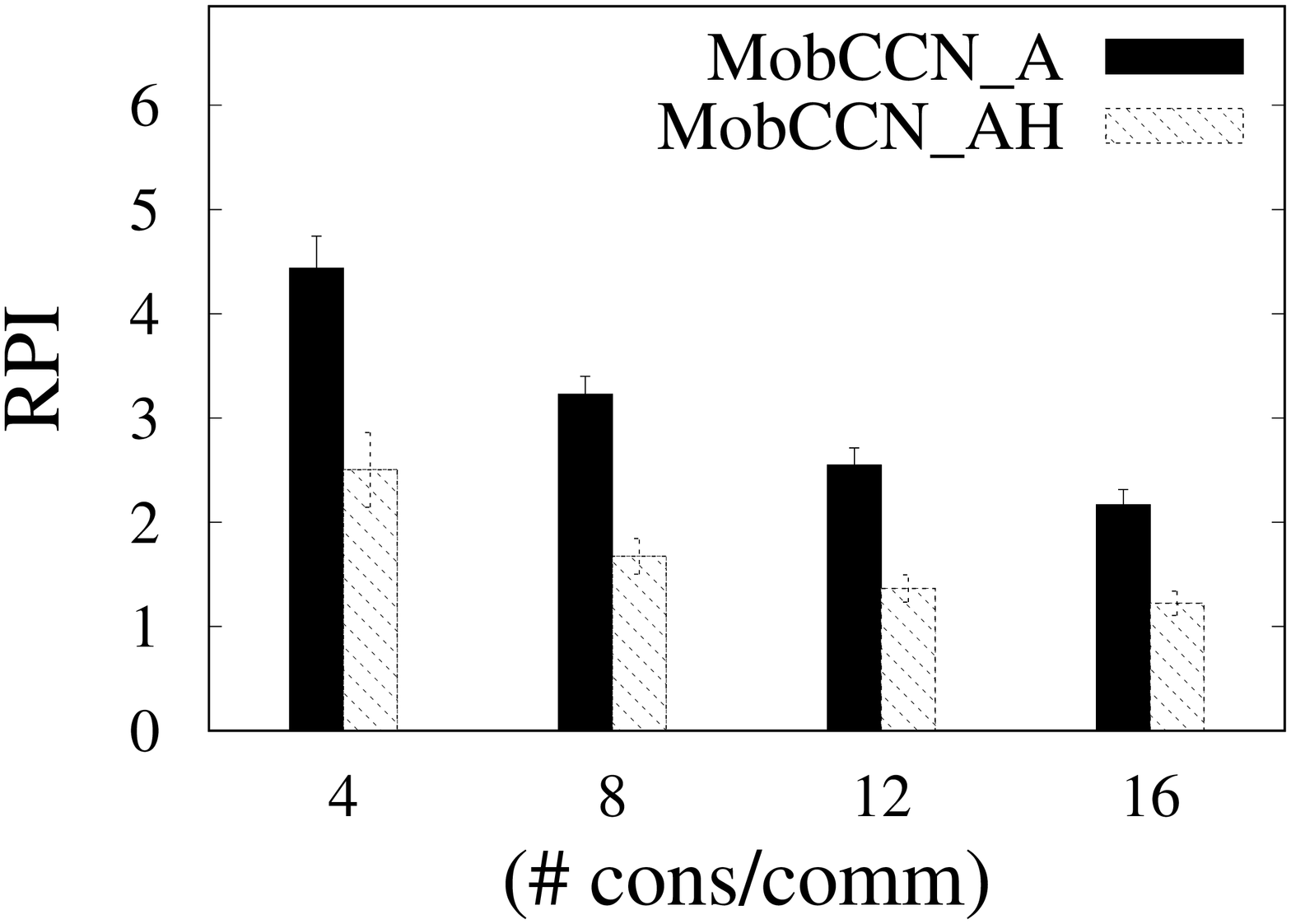}
    }
%
%
    \subfloat[\label{fig:retransmission-metrics:RPD}]{%
      \includegraphics[trim={0cm 0.5cm 0cm 0.6cm},clip,angle=-90,width=0.255\textwidth]{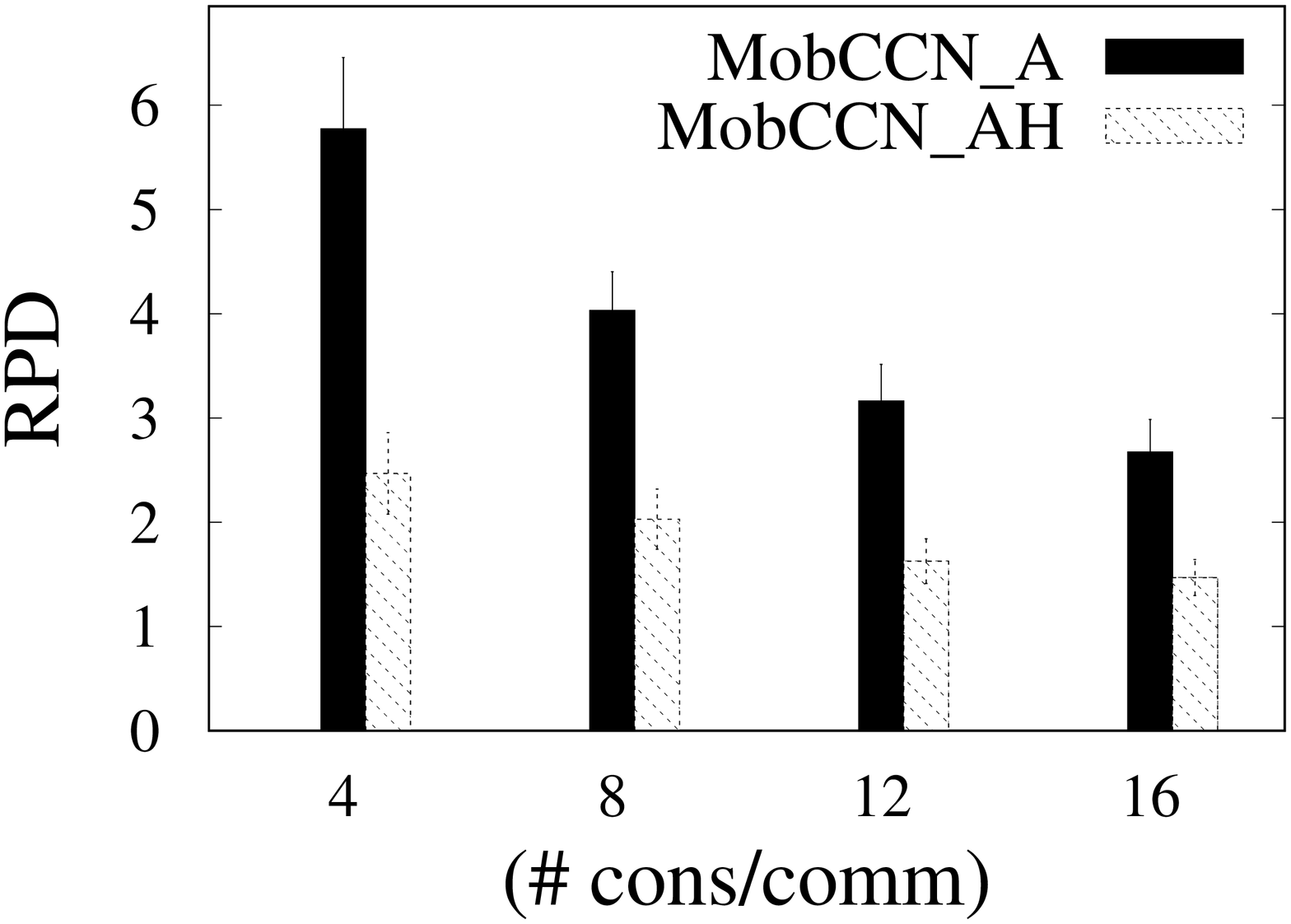}
      }
\vspace{-0.3cm}
\caption{Retransmission performance: retransmission per Interest (RPI) (\ref{fig:retransmission-metrics:RPI}), and retransmission per Data (RPD) (\ref{fig:retransmission-metrics:RPD}) for different consumers/community configurations.}
\label{fig:retransmission-metrics}
\end{figure}
%

%
\subsection{Retransmission performance}
\label{sub:retransmission-performance}
\noindent
In this section we analyse more in detail the properties of the retransmission procedures implemented by MobCCN\_A and \\MoBCCN\_AH, being the two MobCCN variants that maximize the delivery rate, thus guaranteeing the highest reliability performance. To this end, we introduce to new metrics to assess the efficiency of the retransmissions process: namely retransmission per Interest (RPI) and retransmission per Data (RPD). We focus on a scenario with a varying number of consumers per social community, in the range from 4 to 16 consumers. 

Figure \ref{fig:retransmission-metrics:RPI} shows the RPI value for the evaluated scenarios, and we observe that MobCCN\_AH significantly outperforms MobCCN\_A in all the four configurations we considered. Specifically, the number of retransmissions for each Interest generated by the consumers is very limited, with at most 2 retransmissions in the case of 4 consumers per community. On the contrary, MobCCN\_A needs around 5 retransmissions per Interest in the same scenario. Another observation is that both protocols reduce the number of retransmissions as the number of consumers increases. This is another indication of the multiplexing effect achieved by a content centric data-dissemination protocol when the number of interested nodes increases.
As far as RPD (see Figure \ref{fig:retransmission-metrics:RPD}), we can draw similar considerations, and MobCCN\_AH outperforms MobCCN\_A with a significant drop of the number of retransmissions per Data received, which is estimated between 40\% and 58\%. For instance, in the worst case MobCCN\_AH generates less than 3 retransmissions per each successfully received Data packet, while MobCCN\_A requires more that 6 retransmissions for the same outcome. Also for this metric we observe that it is inversely proportional to the number of consumers. 

From the above discussion, we can conclude that another important advantage of MobCCN\_AH is that not only it achieves the highest performance compared to the other retransmission mechanisms that we have explored, but it also reaches such performance by spending less in terms of retransmissions, and consequently limiting further the network resource consumption.

 \begin{figure}[t]
	\hspace{-5mm}
  	\centering
  	\includegraphics[trim={2cm 0.5cm 0cm 0.5cm}, angle=-90,width=0.52\textwidth]{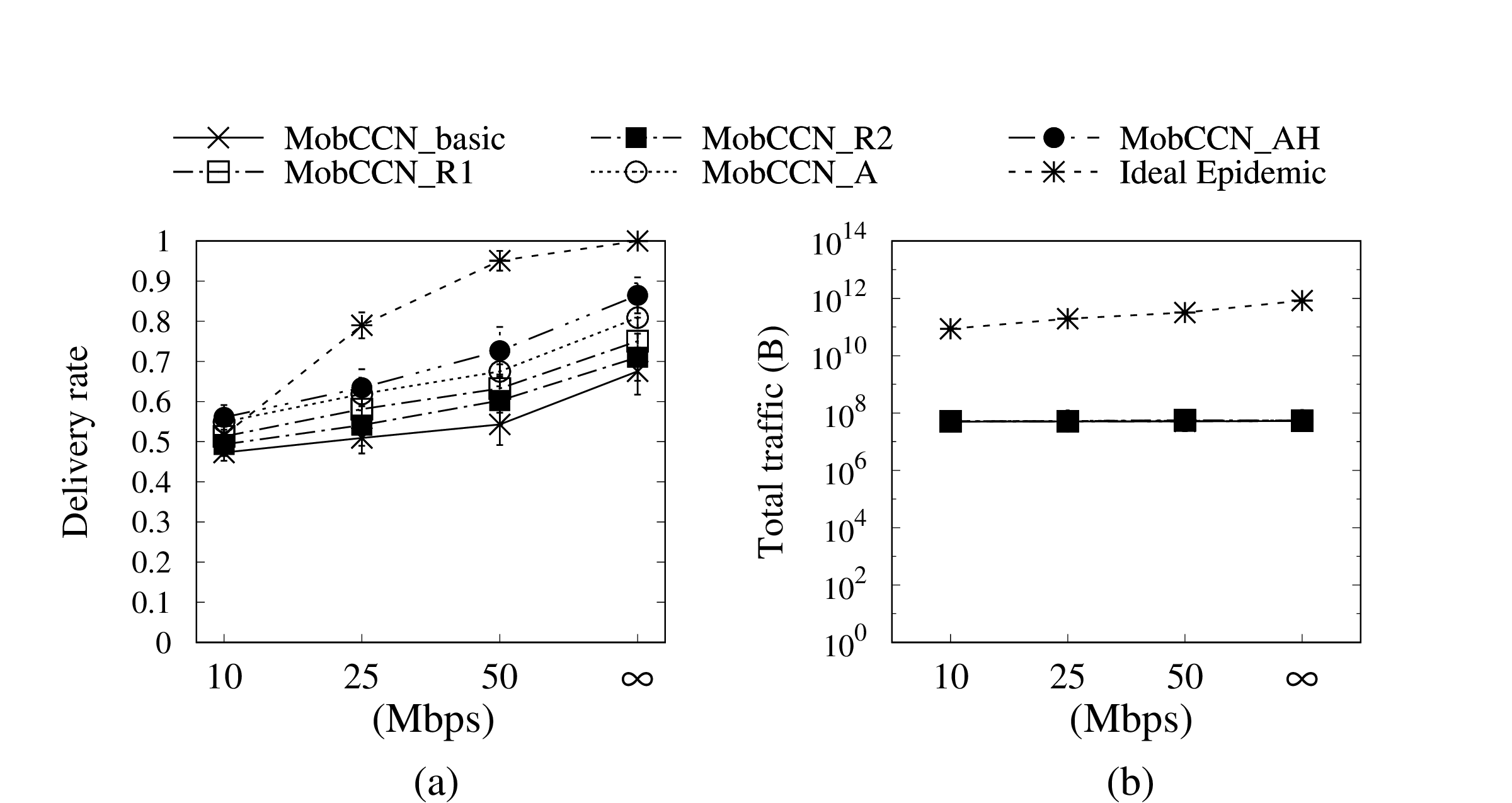}

 	\caption{Delivery rate (\ref{fig:limBW-orig}a) and total traffic (\ref{fig:limBW-orig}b) when varying the available bandwidth.}

 	\label{fig:limBW-orig}
 \end{figure}
%
%
\vspace{-0.2cm}
 \subsection{Impact of bandwidth}
\label{sub:lim-bw-performance}
\noindent
The results shown so far are obtained by assuming the bandwidth of D2D links is infinite, namely all the chunks stored in the node can be transmitted instantaneously and without losses. In this section, we complete our performance analysis by removing such simplifying assumption. More precisely, we assume that the bandwidth of the wireless channel is a fixed and constant value, and that a perfect channel access scheme is used that allows equally sharing the channel bandwidth among the nodes that are in contact. In the following tests we set the size of each chunk equal to $2MB$, and we vary the channel bandwidth from 10 Mbps to 50 Mbps.

\begin{figure}[t]
  	\centering
  	\includegraphics[trim={2cm 0.5cm 0cm 0.5cm}, angle=-90,width=0.53\textwidth]{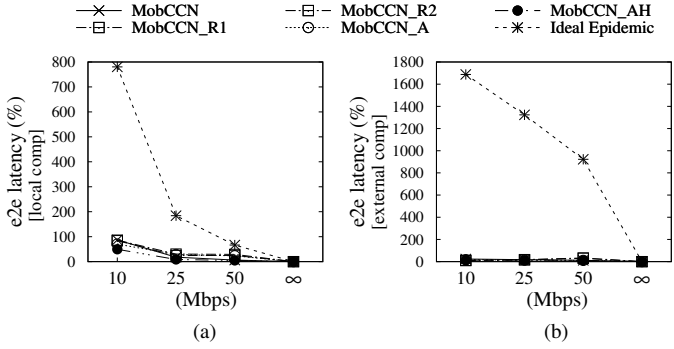}

 	\caption{Breakdown of the end-to-end latency when varying the available bandwidth: percentage increase of the end-to-end latency with respect to the infinite bandwidth case for the local component (\ref{fig:e2e_limBW-orig}a) and external component (\ref{fig:e2e_limBW-orig}b).}

 	\label{fig:e2e_limBW-orig}
 \end{figure}

Looking at the delivery rates (see Figure \ref{fig:limBW-orig}a), we observe that reducing the channel bandwidth results in an overall decrease of the delivery rate for all the protocols, as expected. Two main reasons can explain this result. First of all, the more the neighbours, the lower the available per-node communication bandwidth. Second, a contact may not be sufficient for transmitting all the available contents stored on the node, and a transmitter should wait for the next contact with the same node to complete the transmission. The behaviour of the delivery rate is similar for all the MobCCN variants, and the curves tend to converge to values in the range [0.45-0.55] with 10 Mbps bandwidth. MobCCN\_AH still achieves the highest (but reduced) delivery rate, about 10\% higher than the one of MobCCN\_basic in the worst case with 10 Mbps bandwidth. Delivery rate degrades also for Ideal Epidemic, down to 0.49 in the 10 Mbps case, thus losing its advantage over the MobCCN variants. This behavior is justified by the huge amount of sent traffic (around 775 GB as highlighted by Figure \ref{fig:limBW-orig}b) that excessively consumes the available bandwidth. On the contrary, this effect is less marked with the MobCCN variants since the volume of sent traffic is significantly lower, being stable around 50 MB. As far as the e2e delays, instead of reporting their absolute values, we calculate the percentage increase with respect to the reference value of the infinite bandwidth, splitting them into the local and the external components. As shown in Figure \ref{fig:e2e_limBW-orig}, the bandwidth constraint has a less noticeable impact on the delay performance for the MobCCN variants, and delay increases range from 25\% to 80\%. On the contrary, the bandwidth limitation has a more negative effect for Ideal Epidemic, as e2e delay increases almost 1800\% with respect to the reference value in the 10 Mbps case.

Interestingly, the above results have shown that a bandwidth limitation have a small impact on the e2e delay. This implicitly suggests that the main reasons for the very high delays depend on mobility patterns, namely the very low frequency with which encounters happen and low network density. To confirm this intuition, we also investigate a more dynamic scenario where the inter-contact times between the nodes are considerably reduced. For this purpose, we have manipulated the mobility traces used in the previous tests by setting the inter-contact times to 10\% of the initial ones. Then, in addition to modifying the content request patterns to be in line with the mobility trace obtained, we have performed a new sensitivity evaluation of the protocol parameters since the nodes meet more frequently and set them accordingly, i.e., $ReTX_{threshold}=0$, $T_{ageing}=1800$, and $Hyst=35\%$. As before, we evaluate the system performance under bandwidth constraint assumption.
 \begin{figure}[t]
  	\centering
  	\includegraphics[trim={2cm 0.5cm 0cm 0.5cm}, angle=-90,width=0.53\textwidth]{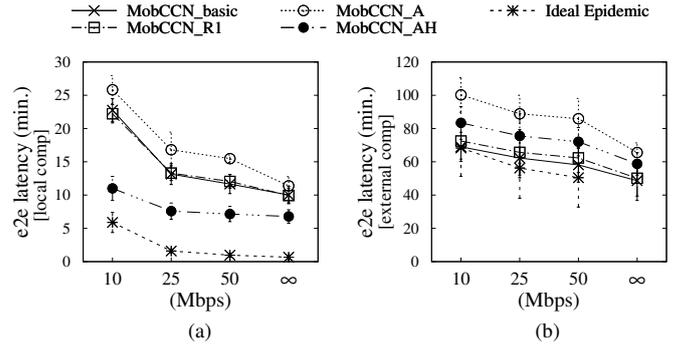}

 	\caption{Breakdown of the end-to-end latency in the scenario with reduced inter-contact times: local component (\ref{fig:e2e_limBW-red}a) and external component (\ref{fig:e2e_limBW-red}b) when varying the available bandwidth.}

 	\label{fig:e2e_limBW-red}
 \end{figure}
 \begin{figure}[t]
  	\centering
  	\includegraphics[trim={2cm 0.5cm 0cm 0.5cm}, angle=-90,width=0.53\textwidth]{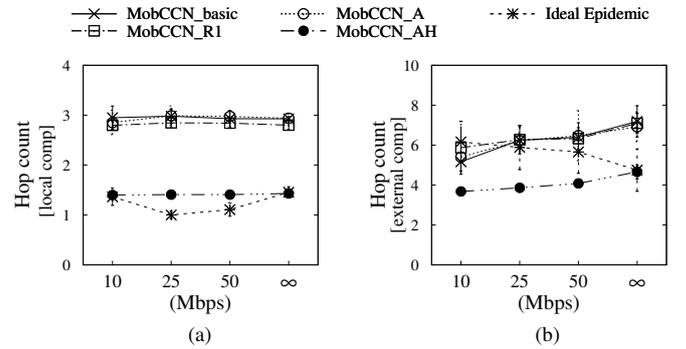}

 	\caption{Breakdown of the hop count in the scenario with reduced inter-contact times: local component (\ref{fig:nHop_limBW-red}a) and external component (\ref{fig:nHop_limBW-red}b) when varying the available bandwidth.}

 	\label{fig:nHop_limBW-red}
 \end{figure}
As far as the delivery rates, the results are aligned with the ones shown in Figure \ref{fig:limBW-orig}a and are omitted. Comparing the latencies (see Figure \ref{fig:e2e_limBW-red}), we observe that: {\it i)} reducing the available bandwidth results in an increase of the e2e delays, but {\it ii)} the latter are significantly reduced both for local requests and for the external ones with respect to the previous set of experiments. The more frequent are the node encounters, the more densely connected is the network and the higher is the latency decrease. Again, Ideal Epidemic experiences the lowest values in both cases, while MobCCN\_AH performs better among the MobCCN family in the local component. This does not hold for the external component where, on the contrary, MobCCN\_AH achieves an intermediate performance while MobCCN\_basic results are the best ones. However, such result is statistically biased as the values over which the delays are averaged are not comparable in the two cases as MobCCN\_basic satisfies only a very low number of external requests. Finally, the hop count metric is shown in Figure \ref{fig:nHop_limBW-red}. As far as the local component (left graph), we observe that path lengths in MobCCN\_AH are comparable with those in Ideal Epidemic with values of 1.5 on average. The path lengths of the remaining MobCCN variants are longer with average values of 3. As far as the external component (see Figure \ref{fig:nHop_limBW-red}b), once again it emerges the MobCC\_AH ability to choose shorter paths, not only with respect to the other MobCCN variants, but also with respect to the Ideal Epidemic. Therefore, postponing the Interest retransmission is advantageous as it allows the nodes to find a more reliable forwarder, namely the traveller.

\section{Conclusions}
\label{sec:conclusions}
\noindent
In this paper, we have investigated the efficiency of different retransmission approaches to improve the reliability of
ICN-based networking protocols for IoT environments. To this end, we have extended the reliability of the data retrieval process of MobCCN, an ICN-based data-centric protocol that we proposed in \cite{Borgia:2016:MCP:2979683.2979695, BORGIA201881} for a mixed IoT environment composed by static and mobile nodes that meet opportunistically and implement ICN functions. This study focused on the analysis of two different classes of retransmission mechanisms. The first class relies on the number of received Interests for the same content to activate the retransmission, while the second relies on the periodic retransmissions of pending Interests. Within each class we then have explored two variants. In the first class, we have considered a single-path approach and a multi-path approach. In the second class, we have studied a retransmission solution based on timeout expiration and one that combines it with a hysteresis mechanism. 
  
Our extensive simulations show that a suitable retransmission strategy is able to improve the reliability of MobCCN while further reducing the network resource consumption. Specifically, the retransmission schemes that employ periodic retransmissions of pending Interests outperform the original MobCCN as well as the retransmission schemes based only the number of received Interests. 
In particular, the solution that combines both periodic retransmissions and a hysteresis mechanism so as to delay a retransmission if the available next-hop forwarder provides only a marginal utility gain, called MobCCN\_AH, achieves the best performance. Indeed, the combination of periodic retransmissions with a careful selection of the best forwarder on the basis of the utility difference among nodes, guarantees that transmissions/retransmissions take place only if needed, that is, if there exists an effective advantage in terms of utility to find the desired content.   

In conclusion, the major advantages of MobCCN\_AH  are: {\it i)} average delivery rates in the order of [85-95]\% in case of infinite bandwidth, with a significant improvement of the number of contents retrieved from external communities (up to 60\%), and average delivery rates of 70\% with 50 Mbps bandwidth; {\it ii)} a 50\% reduction with respect to other MobCCN variants of the length of the network path that are traversed by Data packets, closely approaching the average path lengths that are observed with an Ideal Epidemic protocol; {\it iii)} limited number of retransmissions issued per Interest generated and per Data received. The price to pay is the overhead due to the proactive routing traffic required to maintain updated FIBs, which, however, remains reasonably sized and never exceeds 50MB. Furthermore, this traffic component becomes marginal as soon as the data traffic is more intensive. 
Another downside is a lower delivery rate performance in cases of channels with bandwidth lower than 25Mbps. In order to guarantee higher delivery rates even when the channel is congested, and hence making MobCCN\_AH more flexible and versatile to the various wireless transmission protocols, additional strategies are required. A possible solution is to study adaptive retransmission mechanisms that better follow the evolution of the network by changing for example the timeout parameter.
As future works, we will investigate in-networking caching techniques combined with the MobCCN\_AH routing process to further reduce latencies in retrieving contents.

\section*{Funding}
\noindent
This work is partly funded by EC under the H2020 INFRADEV-2019-3 SLICES-DS (951850), the H2020 INFRAIA-02-2020 SLICES-SC (101008468) and the H2020 ICT-2020-1 MARVEL (957337) projects.


\balance
\typeout{}
\bibliography{rel_MobCCN-FGCS-R2}

\end{document}